\documentclass[12pt]{article}
\usepackage{amssymb,amsfonts,amsmath,latexsym}
\usepackage{color,float,fancyhdr}
\usepackage{graphicx}
\usepackage[latin1]{inputenc}
\usepackage{pstricks,pst-node,pst-text,pst-3d}

\def\eqalign#1{\null\,\vcenter{\openup\jot \m@th
  \ialign{\strut\hfil$\displaystyle{##}$&$
     \displaystyle{{}##}$\hfil \crcr#1\crcr}}\,}

\def\eqalignno#1{\displ@y \tabskip=\centering
  \halign to\displaywidth{\hfil$\@lign\displaystyle{##}$
    \tabskip=0pt &$\@lign\displaystyle{{}##}$
     \hfil\tabskip=\centering
     &\llap{$\@lign##$}\tabskip=0pt\crcr #1\crcr}}

\def\leqalignno#1{\displ@y \tabskip=\centering
  \halign to\displaywidth{\hfil$\@lign\displaystyle{##}$
    \tabskip=0pt &$\@lign\displaystyle{{}##}$
    \hfil\tabskip=\centering &\kern-\displaywith\rlap{$\@lign##$}
    \tabskip=\displaywith\crcr #1\crcr}}

\def\negthickspace{\kern-0.277778em }
\newtheorem{theo}{\sc Theorem}[section]
\newtheorem{prop}{\sc Proposition}[section]

\newtheorem{lem}{\sc Lemma}[section]

\def\R{\mathbb{R}}
\def\N{\mathbb{N}}

\def\E{\mathbb{E}}
\def\P{\mathbb{P}}
\newcommand{\1}{\mbox{\rm 1\hspace{-0.3em}I}}

\def\a{\alpha}
\def\b{\beta}
\def\g{\gamma}

\def\d{\displaystyle}
\def\n{\noindent }

\title{There is a VaR beyond usual approximations}

 \author{M. Kratz \thanks{Marie Kratz, ESSEC Business School, CREAR risk research center, avenue Bernard Hirsch BP 50105, 95021 Cergy-Pontoise Cedex, France; Email: kratz@essec.edu} ~\thanks {Marie Kratz is also member of MAP5, UMR 8145, Univ. Paris Descartes, France} }

\begin{document}

\date{}

\maketitle

 \begin{abstract}
\noindent Basel II and Solvency 2 both use the Value-at-Risk (VaR) as the risk measure to compute the Capital Requirements. In practice, to calibrate the VaR,
a normal approximation is often chosen for the unknown distribution of the yearly log returns of financial assets. This is usually justified by the use of the Central Limit Theorem (CLT), when assuming aggregation of independent and identically distributed (iid) observations in the portfolio model.
Such a choice of modeling, in particular using light tail distributions, has proven during the crisis of 2008/2009 to be an inadequate approximation when dealing with the presence of extreme returns; as a consequence, it leads to a gross underestimation of the risks.\\
The main objective of our study is to obtain the most accurate evaluations of the aggregated risks distribution and risk measures when working on financial or insurance data under the presence of heavy tail and to provide practical solutions for accurately estimating high quantiles of aggregated risks.
We explore a new method, called Normex, to handle this problem numerically as well as theoretically, based on properties of upper order statistics. Normex provides accurate results, only weakly dependent upon the sample size and the tail index. We compare it with existing methods.    
 \end{abstract}
 
\noindent \footnotetext{
\emph{2010 AMSC}: 60F05; 62G32; 62G30; 62P05; 62G20; 91B30; 91G70 
}

\newpage
\section*{Introduction}

\noindent Financial institutions, as banks and insurances, always consider portfolios of individual risks to evaluate their risk exposure. It is the reason why aggregated risks, modeled with random variables (rv's), constitute the basis of internal models developed in those institutions, and are the focus of much investigation to evaluate at best their resulting distribution.\\
In practice, when working on market risk data, e.g. on returns of financial assets which are known to be heavy tailed, it appears that the distribution of the yearly log returns  of financial assets is often approximated by a normal distribution (via  the Central Limit  Theorem (CLT)), assuming the aggregation of iid observations in the portfolio model.
Under this last assumption, there are two main drawbacks when using the CLT for moderate heavy tail distributions (e.g. Pareto with a shape parameter larger than 2). The first one is that, if the CLT may apply to the sample mean because of a finite variance, it also provides a normal approximation with a slow rate of convergence; it may be improved when removing extremes from the sample (see e.g. \cite{hahn:mw} and references therein). Moreover, even if we are interested only in the sample mean, samples of  small or moderate sizes will lead to a bad approximation. Improving the approximation would require to ask for the existence of moments of order larger than 2.
A second drawback is that, when working on heavy tailed aggregated data, the tail may clearly appear, e.g. on QQ-plots, as heavy on high frequency data (e.g. daily ones)  but may become not visible anymore when aggregating them in e.g. monthly or yearly data (i.e. short samples), although it is well known that the tail index of the underlying distribution remains constant under aggregation. The figures on the S\&P 500 returns illustrate very clearly this last issue, as we can see:\\
\begin{minipage}{0.45\linewidth}
\centering
\includegraphics[width=8cm,height=8cm]{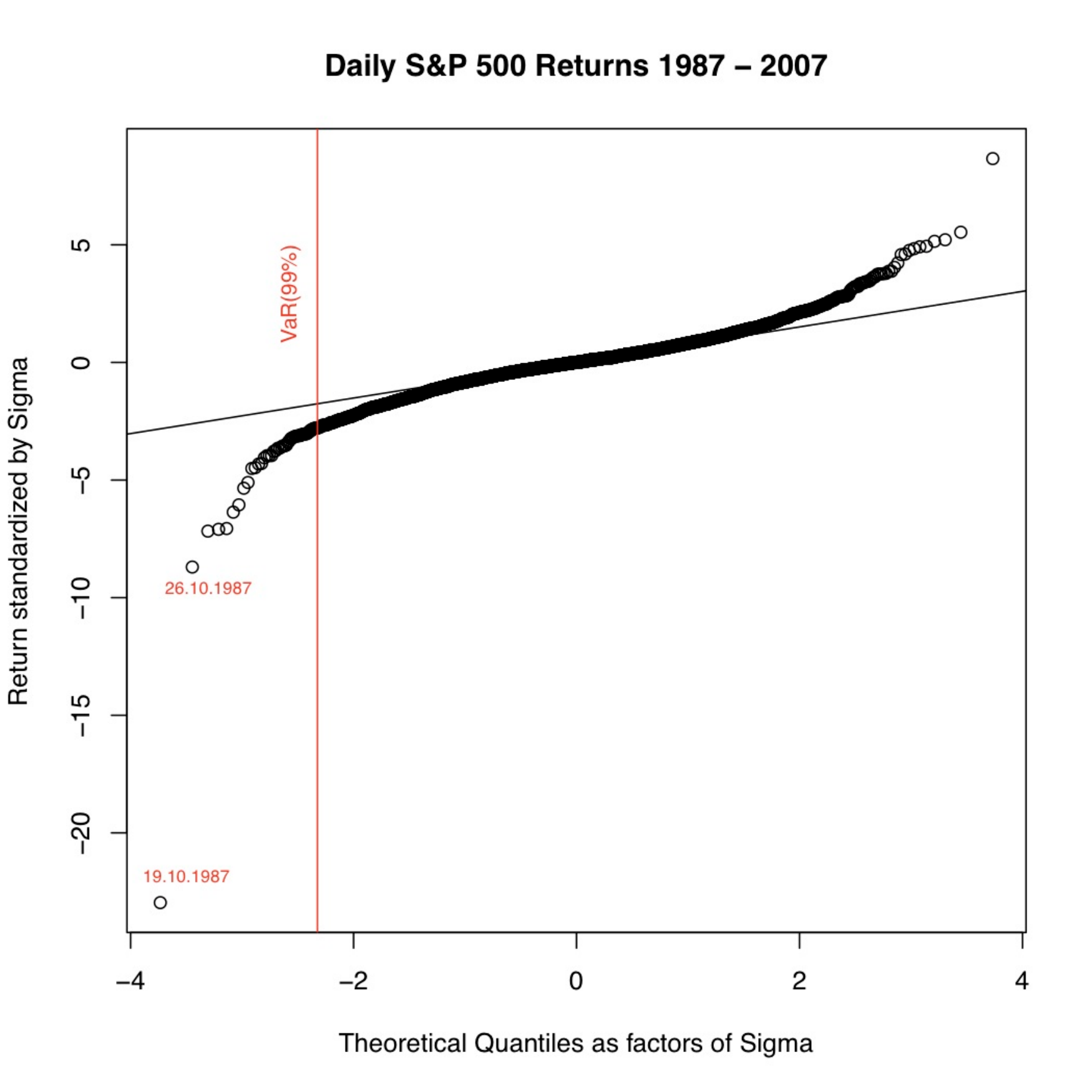} 
\end{minipage}
\hfill
\begin{minipage}{0.50\linewidth}
\centering
\includegraphics[width=8cm,height=8cm]{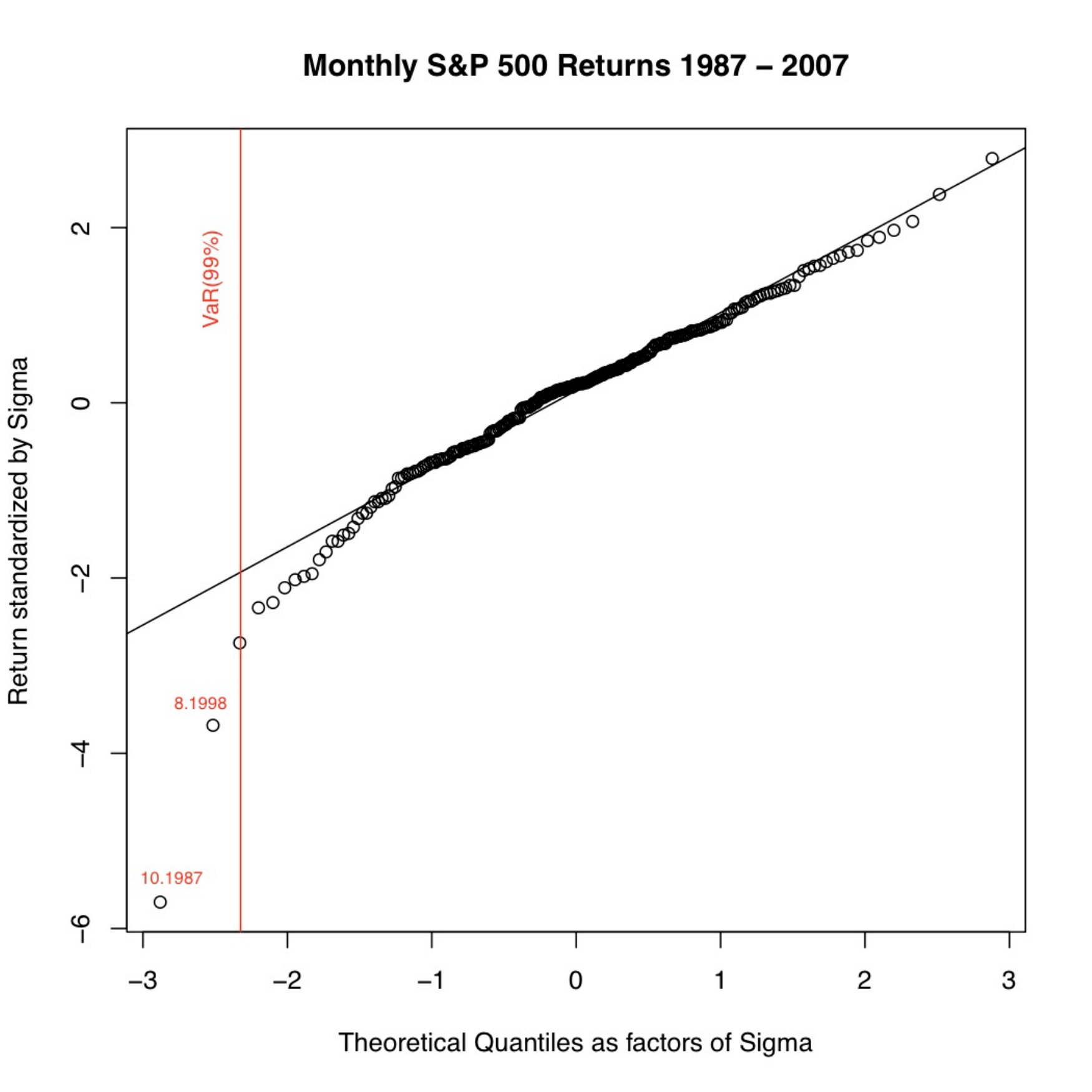} 
\end{minipage}

\n The main objective of this study is to build the most accurate approximation for the distribution of aggregated risks when working on financial data under the presence of fat tail, not only for the mean but also for the tail behavior in order to obtain sharp evaluation of risk measures, and independently of the way we aggregate the risks.  \\
The first section will briefly review the existing methods, from the General Central Limit  Theorem (GCLT) to Extreme Value Theory (EVT) methods, then present some properties of order statistics, known and new ones, useful to construct our method named {\it Normex}. This new approach, inspired by the work of Zaliapin {\it et al.}'s (see  \cite{zaliapin}), will be described in the next section and compared analytically with standard methods. Finally we will apply on simulated samples the different methods, existing ones and Normex,  to compute extreme quantiles which are used as risk measures in solvency calculation. We will then compare numerically their accuracy.

\n With financial/actuarial applications in mind, we use power law models, such as Pareto,  for the marginal distributions of the risks. Note that this is also justified by the Extreme Value Theory (EVT).


\section{A brief review of existing methods and properties of order statistics}

Limit theorems for the sum of independent and identically distributed (iid) random variables (rv's) are well known. Nevertheless, they can be misused in practice and, as a consequence, lead to wrong results when applied to evaluate risk measures for aggregated risks. To help practitioners to be sensitive to this issue, we consider the simple example of aggregated heavy-tailed risks, where the risks are represented by iid Pareto rv's, which is a natural frame for this study, as we are going to see.  \\

\n To aggregate risks implies a decrease of the sample size (of the aggregated risks), hence comes the question of how reasonable it is  to use a limit distribution as an approximation of the true distribution of the aggregated risks, and which type of approximation can be used for any sample size. First we review the existing methods to approximate the distribution of the Pareto sum, from the General Central Limit  Theorem (GCLT) to Extreme Value Theory (EVT) approaches.\\

\n Let us start with some notation.\\

\n $[x]$ will denote the integer part of any non negative real $x$ such that $[x]\le x < [x]+1$.

\n Let $(\Omega, {\cal A},\P)$ be the probability space on which we will be working.

\n Let $\Phi$ and $\varphi$ denote, respectively,  the cumulative distribution function (cdf) and the probability density function (pdf) of the standard normal distribution ${\cal N}(0,1)$, and  $\Phi_{\mu,\sigma^2}$ and $\varphi_{\mu,\sigma^2}$  the cdf and pdf of the normal distribution ${\cal N}(\mu,\sigma^2)$ with mean $\mu$ and variance $\sigma^2$.

\n Let $X$ be a random variable (r.v.), Pareto (type I)  distributed with shape parameter $\a$ and cdf $F$ defined by 
\begin{equation}\label{dfPareto}
\overline F(x):=1-F(x)=x^{-\a}, \quad \a>0, ~x\ge1
\end{equation}
and pdf denoted by $f$.\\ 
Note that the inverse function $F^{\leftarrow}$ of $F$ is given by 
\begin{equation}\label{df-InversePareto}
 F^{\leftarrow}(z)= (1-z)^{-1/\a~}, \quad\text{for}~ 0<z<1
 \end{equation}
Recall that for $\a>1$, $\d \E(X)=\frac{\a}{\a-1}$  and for $\a>2$, $\d var(X)= \frac{\a}{(\a-1)^2(\a-2)}$.\\
We denote by $S_n$ the Pareto sum $\d S_n:=\sum_{i=1}^n X_i$ , $(X_i, i=1,\ldots,n)$ being an $n$-sample with parent r.v. $X$ and associated order statistics $X_{(1)}\leqslant\cdots\leqslant X_{(n)}$.\\

\n We will consider iid Pareto rv's  in this study. Why Pareto? it is justified by the EVT.  Indeed recall the Pickands theorem  (see  \cite{pickands} or e.g. \cite{em:km}) proving that for sufficiently high threshold $u$, the General Pareto Distribution (GPD) $G_{\xi,\sigma(u)}$ (with shape parameter $\xi$ and scale parameter $\sigma(u)$)  is a very good approximation to the excess cdf defined by  $F_u(x)=\P[X-u\le x | X>u]$:
$$
F_u(y) \,\, \underset{u \to \infty} {\approx} \,\, G_{\xi,\sigma(u)}\left(y\right)
$$
When considering risks under the presence of heavy tail, it implies that  the extreme risks follow a GPD with a positive shape parameter $\xi>0$. Therefore, since,
for $\xi>0$, $\displaystyle \overline G_{\xi,\sigma(u)}(y)~\underset{y\to\infty}{\sim} ~cy^{-1/\xi}$, for some constant $c>0$, then it is natural and quite general to consider a Pareto distribution for heavy tailed risks. \\
We may also wonder if the i.i.d. condition is not too  restrictive to keep interest to this study. Again the EVT provides an answer to this provocative question. Indeed, it tells us that  the tail index of the aggregated distribution corresponds to the one of the marginal with the heaviest tail, hence does not really depend on whether we consider the dependence or not. Moreover, when focusing on the evaluation of the VaR risk measure, it is also somehow confirmed by a recent paper by Embrechts et al. (see \cite{em:pr})  providing the worse lower and upper bounds of the VaR of aggregated risks, whatever is the dependency. The bounds appear very close in the Pareto case.  

\subsection{Existing approximations for the aggregated risks}

Let us review the existing methods to approximate the distribution of the Pareto sum, from the General Central Limit  Theorem (GCLT) to Extreme Value Theory (EVT) approaches.\\

\n $\bullet$ {\it A GCLT approach} {\small (see e.g. Samorodnitsky {\it et al.} 1994,  Petrov 1995, Zaliapin {\it et al.} 2005, Furrer 2012)\\

\n The distribution of $S_n$ can be approximated by 
\begin{itemize}
\item[-] a stable distribution whenever $0<\alpha<2$ (via the GCLT) 
\item[-]  a standard normal distribution for $\alpha\ge 2$ (via the CLT for $\a>2$; for $\a=2$, it comes back to a normal limit with a variance different from  $var(X)=\infty$):
\end{itemize}
\begin{eqnarray*}
&\text{If}~ 0<\alpha<2 , &  \frac{S_n - b_n }{n^{1/\alpha}C_\alpha} ~\overset{d}{\rightarrow} ~G_\alpha \quad \text{normalized $\alpha$-stable distribution}\\ 
&\text{If}~ \alpha\ge 2 , &  \frac{1}{d_n}\Big(S_n -\frac{n\alpha}{\alpha-1}\Big) ~\overset{d}{\rightarrow}~\Phi 
\end{eqnarray*} }
with 
 \begin{eqnarray*}
b_n &=&\left\{ 
\begin{array}{lll}
0 & \text{if} &0<\a<1\\
\d \frac{\pi n^2}{2}\int_1^\infty \sin\big(\frac{\pi x}{2n}\big) dF(x)~\simeq~n\left(\log n +1-C-\log(2/\pi)\right)& \text{if} & \a=1\\
n~\E(X)=n\a/ (\a-1)& \text{if} & 1<\a<2
\end{array} 
\right.
\end{eqnarray*}
$\qquad \qquad (C=\text{Euler constant}~0.5772)$ \\
\begin{eqnarray*}
C_\a  = \left\{ 
\begin{array}{ll}
\left(\Gamma(1-\a)\cos(\pi\a/2) \right)^{1/\a}~ \text{if} ~\a \neq 1\\
\pi /2 \quad \text{if} ~ \a=1
\end{array} 
\right.
\!\!\!&;&\!\!\!
d_n = \left\{ 
\begin{array}{ll}
\sqrt{n~var(X)}~=~ \sqrt{\frac{n\a}{(\a-1)^2(\a-2)}} \quad \text{if} ~\a >2\\
\inf\Big\{x: \frac{2n\log x}{x^2} \le 1 \Big\} ~ \text{if} ~\a =2
\end{array} 
\right.
\end{eqnarray*}

\vspace{2ex}

\n $\bullet$ {\it An EVT approach}\\

\n Recall the following result.
\begin{lem}\label{tailMax} (see e.g. \cite{feller}, \cite{em:km})\\
Assume that $X_i, i=1,\ldots,n$ are i.i.d.  rv's with cdf $F$ having a regularly varying tail with tail index $\b\ge 0$, then for all $n\ge 1$,
$$
\overline {F^{n\ast }} (x) ~\sim ~n\overline F (x) \quad\text{as}~x\to\infty
$$
which means that the tail of the cdf of the sum of iid rv's is mainly determined by the tail of the cdf of the maximum of these rv's:
$$
\P[S_n>x] ~\sim ~ \P[\max_{1\le i \le n}X_i>x]  \quad\text{as}~x\to\infty
$$
\end{lem}
It applies of course to Pareto rv's.

\vspace{3ex}

\n $\bullet$ {\it  A mixed approach by Zaliapin et al.} for the case $2/3<\a<2$\\

\n The neat idea of this approach is to rewrite the sum of the $X_i$'s as the sum of the order statistics $X_{(i)}$  and to separate it into two terms, one with order statistics having finite variance and the other as the complement:

$$
S_n = \sum_{i=1}^n X_i~ = ~ \sum_{i=1}^{n-2} X_{(i)} ~+~\big( X_{(n-1)}+X_{(n)} \big) 
$$
with  $var(X_{(i)})<\infty$, $i=1,\cdots,n-2$.\\

\n Assuming the independence of the two subsums and using the CLT  for the finite variance sum, the authors obtained the following approximation for the cdf of $S_n$:
$$
P(S_n\le x) \underset{n\to\infty}{\simeq}  P\Big[{\cal N}\big(m_1(\a,n,2), \sigma^2(\a,n,2)\big)\le x\Big] \times  P\Big[X_{(n-1)}+X_{(n)}\le x \Big]
$$
with
\begin{eqnarray*}
m_1(\a,n,2) &= & \sum_{i=1}^{n-2} \frac{n!\Gamma(n-i+1-1/\a)}{(n-i)!\Gamma(n+1-1/\a)} \\
~\!\!\! \!\!\! m_2(\a,n,2) \!\!\!& \!\!\!= \!\!\!& \!\!\! \frac{n!}{\Gamma(n+1-2/\a)} \!\! \left( \sum_{i=1}^{n-2} \frac{\Gamma(n-i+1-2/\a)}{(n-i)!} +
 \right. \\
  && \qquad\qquad \left.  2 \sum_{j=2}^{n-2} \sum_{i=1}^{j-1}  \frac{\Gamma(n-j+1-1/\a) \Gamma(n-i+1-2/\a)}{(n-j)!\Gamma(n-i+1-1/\a)} \right) \\
\sigma^2(\a,n,2) &=& m_2(\a,n,2)-m^2_1(\a,n,2)
\end{eqnarray*}
Compared with the GCLT method, this approach, even if quite crude, provides a better approximation  for the Pareto sum than the GCLT does, and for an arbitrary number of summands, with a higher degree of accuracy. It gives also a better result for the evaluation of the Value-at-Risk than the  GCLT one does, even when using another rough approximation, namely replacing the quantile of the total Pareto sum with the direct summation of the quantiles of each subsum.
Another drawback would be, when considering the case $\alpha>2$, to remain with one sum of all terms with a finite variance, hence in general with a poor or slow normal approximation.\\

\n Before turning to the construction of Normex, let us recall some properties of order statistics (see e.g. \cite{da:na}) and provide new ones (from straightforward computations) that will be needed in the next session.

\subsection{Some properties of order statistics} \label{order-stat}

\n $\bullet$ {\it Distribution of order statistics} (see e.g. \cite{da:na})\\

\n Recall that the pdf $f_{(i)}$ of $X_{(i)}$ ($1\le i \le n$) is given by
\begin{eqnarray*}
f_{(i)}(x) &=& \frac1{B(i,n-i+1)}F^{~ i-1}(x)\big(1-F(x)\big)^{n-i}f(x),\quad\text{with}~ B(a,b)=\frac{(a-1)!(b-1)!}{(a+b-1)!}, ~ a,b\in\N^*
\end{eqnarray*}
that the joint pdf $f_{(n_1)\ldots(n_k)}$ of the order statistics $X_{(n_j)}$, $j=1,\ldots, k$, with $1\le n_1<\ldots<n_k\le n; 1\le k\le n$ is, for $1<x_1\le \cdots \le x_k$
\begin{eqnarray*}
f_{(n_1)\ldots(n_k)}(x_1,\ldots, x_k)&=& 
n!~\frac{(F(x_1))^{n_1-1}}{(n_1-1)!}~f(x_k)~\frac{(1-F(x_k))^{n-n_k}}{(n-n_k)!}~\prod_{j=1}^{k-1}f(x_j)~\frac{\Big(F(x_{j+1})-F(x_j)\Big)^{n_{j+1}-n_j-1}}{\big(n_{j+1}-n_j-1\big)!}
\end{eqnarray*}
It becomes, for successive order statistics, for $1<x_1\le \cdots \le x_j$, for $i\ge 1$, $j\ge 2$, with $i+j\le n$, 
\begin{eqnarray*}
f_{(i+1)\ldots (i+j)}(x_1,\ldots, x_j)&=& \frac{n!}{i!~(n-i-j)!}F^i(x_1)(1-F(x_j))^{n-i-j}\prod_{l=1}^{j}f(x_l) 
\end{eqnarray*}
For  $\a$-Pareto rv's, those pdf of order statistics are expressed as
\begin{eqnarray}\label{pdf1OrderStat}
f_{(i)}(x) =\frac{n!}{(i-1)!(n-i)!}~\a (1-x^{-\a})^{i-1} x^{-\a(n-i+1)-1}
\end{eqnarray}
and, for $1<x_1\le \cdots \le x_k$,
\begin{eqnarray}\label{jointOrderPdf}
f_{(n_1)\ldots(n_k)}(x_1,\ldots, x_k)
&=&  n!~\a^k~\frac{(1-x_1^{-\a})^{n_1-1}}{(n_1-1)!} \frac{x_k^{-\a(n-n_k+1)-1}}{(n-n_k)!} \prod_{j=1}^{k-1} \frac{x_j^{-\a-1}\Big(x_{j}^{-\a}-x_{j+1}^{-\a}\Big)^{n_{j+1}-n_j-1}}{\big(n_{j+1}-n_j-1\big)!} \qquad\quad
\end{eqnarray}
and for successive order statistics, for $1<x_1\le \cdots \le x_j$, for $i\ge 1$, $j\ge 2$, with $i+j\le n$, 
\begin{eqnarray}\label{jointSuccessivepdf}
f_{(i+1)\ldots (i+j)}(x_1,\ldots, x_j)&=& \frac{n!~\a^j}{i!~(n-i-j)!} (1-x_1^{-\a})^i ~x_j^{-\a(n-i-j)}~ \prod_{l=1}^{j} \frac1{x_l^{\a+1}} 
\end{eqnarray}
Moments of $\a$-Pareto order statistics satisfy 
\begin{eqnarray}
\E[X_{(j)}^p]<\infty & \text{iff} & p < \a(n-j+1)  \label{finiteMoment}  \\
\mbox{and}\quad\E[X_{(j)}^p] &=& \frac{n!~\Gamma(n-j+1-p/\a)}{(n-j)!~\Gamma(n+1-p/\a)} \nonumber
\end{eqnarray}
and, for $1\le i<j\le n$,
\begin{eqnarray}
\E[X_{(i)}X_{(j)}]<\infty & \text{iff} & \min\big(n-j+1~, (n-i+1)/2\big) > 1/\a \label{finite2ndMoment} \\
\mbox{and}\quad \E[X_{(i)}X_{(j)}] &=& \frac{n!~\Gamma(n-j+1-1/\a) \Gamma(n-i+1-2/\a)}{(n-j)!~\Gamma(n-i+1-1/\a)\Gamma(n+1-2/\a)} \nonumber\\
~&&~\nonumber
\end{eqnarray}

\newpage
\n $\bullet$ {\it Conditional distribution of order statistics}\\

\n Let us look now at conditional distributions, in the general case (with the notation $f$ and $F$), then for Pareto rv's.\\
We deduce from \eqref{pdf1OrderStat} and \eqref{jointOrderPdf} that the joint pdf of $X_{(i)}$ given $X_{(j)}$, for $1\le i < j \le n$, is,  for $x\le y$,
\begin{eqnarray}\label{x-given-y}
f_{X_{(i)}/(X_{(j)}=y)}(x)&=& \frac{(j-1)!}{(i-1)!(j-i-1)!}f(x)\big(F(y)-F(x)\big)^{j-i-1}~\frac{F^{i-1}(x)}{F^{j-1}(y)}  \nonumber\\
&=& \frac{\a~(j-1)!}{(i-1)!(j-i-1)!}~x^{-\a-1}\big(x^{-\a}-y^{-\a}\big)^{j-i-1}~\frac{(1-x^{-\a})^{i-1}}{(1-y^{-\a})^{j-1}}
\end{eqnarray}
and that the joint pdf of $(X_{(i)},X_{(j)})$ given $X_{(k)}$, for $1\le i < j < k \le n$ is,  for $x\le y\le z$,
\begin{eqnarray}\label{xy-given-z}
f_{X_{(i)},X_{(j)}/(X_{(k)}=z)}(x,y)&=& \frac{(k-1)!}{(i-1)!(j-i-1)!(k-j-1)!}~f(x)f(y)   \nonumber\\
&& \qquad \times\frac{F^{i-1}(x)\big(F(y)-F(x)\big)^{j-i-1}\big(F(z)-F(y)\big)^{k-j-1}}{F^{k-1}(z)}  \nonumber\\
&=& \frac{\a^2~(k-1)! ~x^{-\a-1}y^{-\a-1}~(1-x^{-\a})^{i-1}\big(x^{-\a}-y^{-\a}\big)^{j-i-1}\big(y^{-\a}-z^{-\a}\big)^{k-j-1}}{(i-1)!(j-i-1)!(k-j-1)!~(1-z^{-\a})^{k-1}} \qquad
\end{eqnarray}
Using \eqref{pdf1OrderStat} and \eqref{jointSuccessivepdf} provides, for $y\le x_1\le \ldots\le x_{j-1}$,
\begin{eqnarray}\label{cdtionalSuccessivepdf}
f_{X_{(i+2)}\ldots X_{(i+j)}/X_{(i+1)}=y}(x_1,\ldots, x_{j-1})\!\!\!&=& \!\!\!\frac{(n-i-1)!}{(n-i-j)!~\big(1-F(y)\big)^{n-i-1}}~(1-F(x_{j-1}))^{n-i-j}\prod_{l=1}^{j-1}f(x_l) \nonumber \\ 
&=& \!\!\!\frac{(n-i-1)!~\a^{j-1}}{(n-i-j)!~y^{-\a(n-i-1)}}~\frac1{x_{j-1}^{\a(n-i-j+1)+1}}\prod_{l=1}^{j-2}\frac1{x_l^{\a+1}} 
\end{eqnarray}
Then we can compute the first conditional moments. 
We obtain, using \eqref{x-given-y},
\begin{eqnarray}\label{cdtionalMean}
\E\big[X_{(i)}/X_{(j)}=y\big] &=& \frac{(j-1)!}{(i-1)!(j-i-1)!~F^{j-1}(y)}~\int_1^y x F^{i-1}(x) \big(F(y)-F(x)\big)^{j-i-1} dF(x) \nonumber\\
&=&  \frac{(j-1)!}{(i-1)!(j-i-1)!}\int_0^1 \!\!\! F^{\leftarrow}\Big(uF(y)\Big)~ u^{i-1}(1-u)^{j-i-1} ~du \qquad\\
&& (\text{with the change of variables}~u=F(x)/F(y) ) \nonumber\\
&=&  \frac{(j-1)!}{(i-1)!(j-i-1)!} \int_0^1 \big(1-uF(y)\big)^{-\frac1\a}u^{i-1}(1-u)^{j-i-1}~du \nonumber \\
&\simeq& \frac{1}{B(i,j-i)} \left\{B(i,j-i)+\sum_{l\ge 1}B(i+l,j-i) \frac{(F(y))^l}{l!}\prod_{m=0}^{l-1}(m+1/\a)
\right\} \nonumber\\
&=& 1+\frac{\Gamma(j)}{\Gamma(i)}\sum_{l\ge 1}\frac{\Gamma(i+l)}{l~\Gamma(j+l)\Gamma(l)} (F(y))^l\prod_{m=0}^{l-1}(m+1/\a)\nonumber
\end{eqnarray}
and, with the same change of variables,
\begin{eqnarray}\label{cdtionalE2}
\E\big[X^2_{(i)}/X_{(j)}=y\big] &=& \frac{(j-1)!}{(i-1)!(j-i-1)!}\int_0^1 \Big[F^{\leftarrow}\Big(uF(y)\Big)\Big]^2~ u^{i-1}(1-u)^{j-i-1} ~du \qquad \\
&=&  \frac{(j-1)!}{(i-1)!(j-i-1)!} \int_0^1 \big(1-uF(y)\big)^{-\frac2\a}u^{i-1}(1-u)^{j-i-1}~du \nonumber \\
&\simeq& 1+\frac{\Gamma(j)}{\Gamma(i)}\sum_{l\ge 1}\frac{\Gamma(i+l)}{l~\Gamma(j+l)\Gamma(l)} (F(y))^l\prod_{m=0}^{l-1}(m+2/\a)\nonumber
\end{eqnarray}
and, for $1\le i<j<k \le n$, via \eqref{xy-given-z},
\begin{eqnarray}\label{cdtionalEij}
\E\big[X_{(i)}X_{(j)}/X_{(k)}=y\big] &=&  \frac{(k-1)!}{(i-1)!(j-i-1)!(k-j-1)!~F^{k-1}(y)} \int_1^y x_i~ F^{i-1}(x_i) \nonumber \\
&&  \Big(\int_{x_i}^y x_j \big(F(x_j)-F(x_i)\big)^{j-i-1}\big(F(y)-F(x_j)\big)^{k-j-1} dF(x_j)\Big)~dF(x_i) \nonumber\\
&=&  \frac{(k-1)!}{(i-1)!(j-i-1)!(k-j-1)!} \int_0^1 F^{\leftarrow}\Big(uF(y)\Big)~ u^{i-1} \nonumber \\
&&  \qquad \Big(\int_u^1 F^{\leftarrow}\Big(vF(y)\Big)~
 \big(v-u\big)^{j-i-1}\big(1-v \big)^{k-j-1} dv\Big)~du  \\
 && \big(\text{with the change of variables}~u=F(x_i)/F(y)~\text{and}~v=F(x_j)/F(y) \big) \nonumber\\
 &=&  \frac{(k-1)!}{(i-1)!(j-i-1)!(k-j-1)!} \int_0^1\big(1-uF(y)\big)^{-\frac1\a}~ u^{i-1} \nonumber \\
&&  \qquad \Big(\int_u^1 \big(1-vF(y)\big)^{-\frac1\a}~
 \big(v-u\big)^{j-i-1}\big(1-v \big)^{k-j-1} dv\Big)~du \nonumber
\end{eqnarray}
where $F(y)=1-y^{-\a}$.\\

\n Moreover, the joint conditional distribution of $(X_{(i+1)},\ldots , X_{(p-1)})$ given $(X_{(k)}=x_k, k\le i, k\ge p)$, for  $1\le i < p \le n$,   denoted by 
$f_{X_{(i+1)},\ldots , X_{(p-1)}~/~(X_{(k)}=x_k,\, k\le i, \, k\ge p)}$, or \\$f_{(i+1),\ldots , (p-1)~/~(X_{(k)}=x_k, \, k\le i, \, k\ge p)}$ when no ambiguity, is, 
for $x_1<\ldots <x_n$,
\begin{eqnarray}
&& f_{(i+1),\ldots , (p-1)~/~(X_{(k)}=x_k, \, k\le i, \, k\ge p)} (x_{i+1},\ldots , x_{p-1}) ~ =  \nonumber \\ 
 && \frac{(p-i-1)!}{\Big(F(x_{p}) - F(x_i)\Big)^{p-i-1}} \prod_{l=i+1}^{p-1} f(x_l)  
 ~ = ~  \frac{(p-i-1)! ~\a^{p-i-1}}{\Big(x_{i}^{-\a} - x_p^{-\a}\Big)^{p-i-1}}~\prod_{l=i+1}^{p-1} \frac1{x_l ^{\a+1}}  \quad\label{conditionalDfOrder}
\end{eqnarray}
It implies that $X_{(i+1)},\ldots , X_{(p-1)}$ are independent of $X_{(1)},\ldots , X_{(i-1)}$ and $X_{(p+1)},\ldots , X_{(n)}$ when $X_{(i)}$ and 
$X_{(p)}$ are given, and that the order statistics form a Markov chain. 
 
\newpage 


\section{Normex : a mixed normal and extremes limit}

\n  In this approach, inspired by Zaliapin {\it et al.}'s paper, we go further in the direction of separating mean and extreme behaviors in order to improve approximations and to settle a method in a formal way. 
It means to answer the question of how many largest order statistics $X_{n-j}, j>k$, would explain the divergence between the underlying distribution and the normal, stable respectively, approximation when considering a Pareto sum with $\a\ge 2$, or $\a<2$ respectively. 
Normex gives an answer in a simple and efficient way.\\
 
\n We are mainly interested in the case of a shape parameter larger than 2, since it is the usual case when studying market risk data, for instance. For such a case, the CLT applies because of the finiteness of the 2nd moment but provides wrong results for the tails, as expected. Indeed, the CLT only concentrates on the average behavior; it is equivalent to the CLT on the trimmed sum ({\it i.e.} $S_n$ minus a given number of the largest order statistics) (see \cite{mori}), which emphasizes that the tail is not considered, and  the rate of convergence improves for trimmed sums (see e.g. \cite{hahn:mw}, \cite{hall}). \\
Moreover, as already mentioned, a fat tail behavior may clearly appear on high frequency data but be not visible anymore (empirically) when aggregating data or when considering short samples, although it is well known that shape parameter of the underlying distribution remains constant under aggregation. Hence we really have to be aware that using CLT to obtain information on something else than the average is simply wrong in presence of fat tails, even if in some situations  the plot of the empirical distribution fits well a normal one.\\

\n Although our focus will be mainly ob the case $\a\ge 2$, we will develop Normex for any $\a\in (0;4]$. We aim at determining in an 'optimal way' (in order to improve at most the distribution approximations) the number $k$ that corresponds to a threshold when splitting  the sum of order statistics into two subsums, the second one constituted by the $k$ largest order statistics, under realistic assumptions. We will drop in particular Zaliapin's et al 's assumption of independence between the two subsums.\\

\n Although the study is developed on the Pareto-example, note that its goal is to propose a method that may be applied to other examples and to real data, hence this choice of looking for limit theorems in order to approximate the true (and most of the time unknown) distribution. \\

\subsection{How to fit for the best mean behavior of aggregated heavy tail distributed risks?}
 
\n Let us start by studying the behavior of the trimmed sum $T_k$ when writing down the sum $S_n$ of the iid $\a$-Pareto rv's (with $\a>0$), $S_n:=\sum_{i=1}^n X_i$, as
\begin{equation}\label{eq:decompSn}
S_n=T_{k}+ U_{n-k}\quad \text{with}\quad T_k:=\sum_{j=1}^{n-k} X_{(j)} \quad \text{and} \quad U_{n-k}:=\sum_{j=0}^{k-1} X_{(n-j)} 
\end{equation}
Much literature, since the 80's, has been concerned with the behavior of trimmed sums by removing extremes from the sample; see e.g. \cite{hall}, \cite{mori}, \cite{hahn:mw}.\\
The main issue is the choice of the threshold $k$, in order to use the CLT but also to improve its fit since we want to approximate the behavior of $T_k$ by a normal one. 

\n We know that a necessary and sufficient condition for the CLT to apply on $T_k$ is to require  the summands $X_{(j)}$, $j=1,\ldots,k$, to be $L^2$-rv's.
But we also know that requiring only the finitude of the 2nd moment may lead to a poor normal approximation, if higher moments do not exist, as occurs for instance with financial market data. 
In particular, including the finitude of the third moment provides a better rate of convergence to the normal distribution in the CLT (Berry-Ess\'een inequality).
\n Another information that might be quite useful to improve the approximation of the distribution of $S_n$ with its limit distribution, is the Fisher index, defined by the ratio $\d \gamma= \frac{\E[(X-\E(X))^4]}{ (var(X))^2}$,  which is a kurtosis index. The skewness  $\d \E[(X-\E(X))^3] / (var(X))^{3/2}$  of $X$ and $(\gamma - 3)$ measure the closeness of the cdf $F$ to $\Phi$. Hence we will choose $k$ based on the condition of existence of the 4th moment of the summands of $T_k$ (i.e. the first $n-k$ order statistics)\\
Note the following Edgeworth expansion involving the Hermite polynomials $(H_n,~n\ge 0)$ which points out that requiring the finitude of the 4th moments appears as  what we call the 'optimal' solution (of course, the higher order moments exist, the finer the normal approximation becomes, but it would imply too strong conditions, difficult to handle).  
If $F_n$ denotes the cdf of the standardized $S_n$ defined by $\d \frac{S_n-n\E(X)}{\sqrt{n~var(X)}}$, then 
\begin{equation}\label{edgeworth}
F_n(x)-\Phi(x) = \frac1{\sqrt{n}}Q_1(x) + \frac1n Q_2(x) + o(1/n)
\end{equation}
uniformly in $x$, with 
\begin{eqnarray*}
Q_1(x)&=& -\varphi(x) \frac{H_2(x)}{6}~\frac{\E[(X-\E(X))^3]}{(var(X))^{3/2}}\\
Q_2(x)&=& -\varphi(x) \Big\{ \frac{H_5(x)}{72}~\frac{\big(\E[(X-\E(X))^3]\big)^2}{(var(X))^3} + 
\frac{H_3(x)}{24}~ \big(\gamma -3 \big) \Big\}\\
\text{and}\quad&&\\
H_2(x)&=& x^2-1; \quad H_3(x)=x^3-3x; \quad  H_5(x)=x^5-10 x^3 + 15 x
\end{eqnarray*}
The rate of convergence appears clearly as $n^{-\delta/2}$ whenever $\E[X^{2+\delta}]<\infty$, $\delta>0$.\\
Note that in our Pareto case, the skewness  $\gamma_1:=\frac{\E[(X-\E(X))^3]}{(var(X))^{3/2}}$ and the excess kurtosis $\gamma_2:=\gamma - 3=\frac{\E[(X-\E(X))^4]}{ (var(X))^2}-3$  are, respectively,
$$
\begin{array}{lcl}
\gamma_1=\frac{2(1+\a)}{\a-3}\sqrt{\frac{\a-2}{\a}} & \mbox{if} & \a>3\\
\\
\gamma_2=\frac{6(\a^3+\a^2-6\a-2)}{\a(\a-3)(\a-4)} & \mbox{if} & \a>4
\end{array}
$$

\n Therefore we set $p=4$ (but prefer to keep the notation $p$ so that it remains general) to obtain what we call an 'optimal' approximation. Then we select the threshold $k=k(\a)$ such that
\begin{equation}\label{momentp}
\E(X^p_{(j)})
\left\{ 
\begin{tabular}{c}
 $< \infty \quad\forall j\le n-k$\\
$=\infty \quad \forall j> n-k$\\
\end{tabular}
\right.
\end{equation}
which applied to our case of $\a$-Pareto iid rv's, using \eqref{finiteMoment}, gives:
\begin{equation}\label{moment4}
k>\frac{p}{\a} -1
\end{equation}
This condition allows then to determine a fixed number $k=k(\a)$ as a function of the shape parameter $\a$ of the underlying heavy tailed distribution of the $X_i$'s  
but not of the size $n$ of the sample. We can take it as small as possible in order to fit for the best both the mean and tail behaviors  of $S_n$. Note that we look for the smallest possible $k$ to be able to compute explicitly the distribution of the last upper order statistics appearing as the summands of the second sum $U_{n-k}$.
For this reason, based on condition \eqref{moment4}, we will choose 
\begin{equation}\label{k-mixedMethod}
k=[p/\a -1]+1
\end{equation}

\n Let us summarize in the table below the necessary and sufficient condition on $\a$ to have the existence of the $p$-th moments for the upper order statistics, for $p=2,3,4$ respectively,(and for $\a>1/4$; we could of course complete the table for any choice of $\a>0$ using \eqref{finiteMoment}), using \eqref{moment4} written as $ \d \a>\frac{p}{k+1}$.\\

\begin{table}[h]
\begin{center}
\begin{tabular}{|c||c|c|c|c|}
\hline
$k$ &  $\E(X^p_{(n-k)})$ & $p=2$ & $p=3$ & $p=4$  \\
\hline\hline 
0 & $\E(X^p_{(n)})<\infty$ &  iff  $\a>2$ &  iff $\a>3$ &  iff  $\a>4$ \\
\hline
1 & $\E(X^p_{(n-1)})<\infty$ &  iff  $\a>1$ &  iff  $\a>3/2$ &  iff  $\a>2$ \\ 
\hline 
2 & $\E(X^p_{(n-2)})<\infty$ &  iff  $\a>2/3$ &  iff  $\a>1$ &  iff  $\a>4/3$  \\
\hline
3 & $\E(X^p_{(n-3)})<\infty$ &  iff  $\a>1/2$ &  iff  $\a>3/4$ &  iff  $\a>1$ \\
\hline
4 & $\E(X^p_{(n-4)})<\infty$ &  iff $\a>2/5$ &  iff  $\a>3/5$ &  iff  $\a>4/5$ \\
\hline
5 & $\E(X^p_{(n-5)})<\infty$ &  iff  $\a>1/3$ &  iff  $\a>1/2$ &  iff  $\a>2/3$ \\
\hline
6 & $\E(X^p_{(n-6)})<\infty$ &  iff  $\a>2/7$ &  iff  $\a>3/7$ &  iff  $\a>4/7$ \\
\hline
7 & $\E(X^p_{(n-7)})<\infty$ &  iff  $\a>1/4$ &  iff  $\a>3/8$ &  iff  $\a>1/2$ \\
\hline
\end{tabular}
\caption{\sf Necessary and sufficient condition on $\a$ for having $\E(|X_{(n-k)}|^p)<\infty$ }
\end{center}
\end{table}

\n We deduce the value of the threshold $k=k(\a)$ satisfying \eqref{moment4} for which the 4th moment is finite, according to the set of definition of $\a$:
\begin{table}[h]
\begin{center}
 \begin{tabular}{|c|c|c|c|c|c|c|c|}
 \hline
$\a \in I(k)$ with $I(k)=$  & $]\frac12; \frac47]$ & $]\frac47; \frac23]$ & $]\frac23; \frac45]$ & $]\frac45; 1$ & $] 1; \frac43]$ & $]\frac43; 2[$ & [2,4]\\
\hline 
$k=k(\a)$ =  & 7 & 6 & 5 & 4 & 3 & 2 & 1\\ 
\hline
 \end{tabular}
  \caption{\sf Value of $k(\a)$  for having up to $\E(|X_{(n-k(\a))}|^4)<\infty$}
 \end{center}
 \end{table}

\n We notice from this table that we would use Zaliapin {\it et al.}'s decomposition $S_n= \sum_{j=1}^{n-2} X_{(j)} +\sum_{j=0}^{1} X_{(n-j)} $ only when $\a\in ]\frac43; 2[$, using then the limit distribution of each term  (a normal one for the first sum and the exact joint distribution of the two largest observations for the second one) to approximate the distribution of $S_n$.  When considering, as they do, $\a>2/3$, we would rather introduce the decomposition  $S_n= \sum_{j=1}^{n-k} X_{(j)} +\sum_{j=0}^{k-1} X_{(n-j)} $, with $k$ varying from 2 to 5 depending on the value of $\a$, to improve the approximation of the distribution of $S_n$.

\subsection {A conditional decomposition}

\n Whatever is the size of the sample, because of the handy magnitude of $k$, we are able to compute explicitly the distribution of the last upper order statistics appearing as the summands of the second sum $U_{n-k}$ defined in \eqref{eq:decompSn}. The choice of $k$ allows also to obtain a good normal approximation for the distribution of the trimmed sum $T_k$. Nevertheless, since $T_k$  and $U_{n-k}$ are not independent, we decompose the Pareto sum $S_n$ in a slightly different way than in  \eqref{eq:decompSn} (but keeping the same notation), namely
\begin{equation}\label{decompSnCdtional}
S_n=T_{k}+ X_{n-k+1}+U_{n-k+1} 
\end{equation}
and use the property of conditional independence (recalled in $\S$\ref{order-stat}) between the two subsums $T_k /X_{(n-k+1)}$ and $U_{n-k+1}/X_{(n-k+1)}$ (for $k\ge 1$).\\
Then we obtain the following approximation of the distribution of $S_n$, for $k\ge 1$ (i.e. when the $p$th moment of the $k$ largest order statistics do not exist).
\begin{theo}\label{dfSn2-cdtional}~
The cdf of $S_n$ expressed in \eqref{decompSnCdtional} with $k=k(\a)\ge 1$ defined in \eqref{k-mixedMethod}, can be approximated by $G_{n,\a,k}$ defined for any $x\ge 1$ by 
\begin{eqnarray*}
G_{n,\a,k}( x) &=&
\left\{\begin{array}{ll}
\d n\a  \int_1^x  \frac1{\sigma(y)} ~\frac{(1-y^{-\a})^{n-1}}{y^{1+\a}}\int_0^{x-y}~ \varphi \left(\frac{v-m_1(y)}{\sigma(y)}\right) dv ~dy  & \text{if }~k=1\\
&\\
\d \int_1^x  f_{(n-k+1)}(y) \int_0^{x-y} \varphi_{m_1(y),\sigma^2(y)}\ast  h_y^{(k-1)\ast } (v)dv ~dy & \text{if }~k\ge 2\\
\end{array}
\right.
\end{eqnarray*}
with $f_{(i)}$ computed in \eqref{pdf1OrderStat},  $h_y$ the probability density function of a Pareto rv with parameters  $\a$ and $y$, {\it i.e.} defined by 
$\displaystyle h_y(x)=\frac{\a~y^\a}{x^{\a+1}}~\1_{(x\ge y)}$,
and where the mean $m_1(y)$ and  the variance $\sigma^2(y)$ of the normal density $\varphi_{m_1(y),\sigma^2(y)}$ are defined respectively by
\begin{eqnarray}\label{m1condi}
m_1(y)=m_1(\a,n,k,y) &=& \frac{n-k(\a)}{1-y^{-\a}} \times \left\{
\begin{array}{lc}
\displaystyle \frac{1-y^{1-\a}}{1-1/\a} & \text{if}~\a\neq 1\\
&\\
\ln(y) & \text{if}~\a=1
\end{array}
\right.
\end{eqnarray}
and 
\begin{eqnarray}\label{sigma2condi}
\sigma^2(y)&=&\sigma^2(\a,n,k,y):= m_2(\a,n,k,y)-m^2_1{(\a,n,k,y)}
\nonumber\\
&=& (n-k(1))~ y \left( 1-\frac{y\ln^2 (y)}{(y-1)^2}\right) \1_{(\a=1)} ~ + ~  2(n-k(2)) ~\frac{y^2}{y^2-1}\left( \ln(y)-2~\frac{y-1}{y+1}\right) \1_{(\a=2)} \nonumber \\
&& + ~ \frac{n-k(\a)}{1-y^{-\a}} \left(\frac{1-y^{2-\a}}{1-2/\a} -\frac1{(1-1/\a)^2}\times\frac{(1-y^{1-\a})^2}{1-y^{-\a}} \right)\1_{(\a\neq 1,2)} \qquad
\end{eqnarray}
\end{theo}

~\\
\n {\it Comments} 
\begin{enumerate}
\item The distribution $G_{n,\a,k}$ can also be expressed as
\begin{eqnarray}
G_{n,\a,1}(x) &=& \a~ n \int_1^x  \frac{(1-y^{-\a})^{n-1}}{y^{1+\a}} \left(\Phi\Big(\frac{m_1(y)}{\sigma(y)}\Big)-\Phi\Big(\frac{m_1(y)-(x-y)}{\sigma(y)}\Big)\right)~dy \qquad\quad \label{Gbis-k1}\\
\text{and, for}~ k\ge 2,&&  \nonumber  \\
G_{n,\a,k}(x) &=& \int_1^x  \frac{f_{(n-k+1)}(y)}{\sigma(y)}  \int_0^{x-y}\!\!\! \left(\int_0^v\!\!\!  \varphi\Big(\frac{v-u-m_1(y)}{\sigma(y)}\Big) h_y^{(k-1)\ast} (u)du\right)~dv ~dy \qquad\quad \label{Gbis-k2}
\end{eqnarray} 
\item Note that we considered iid Pareto rv's 
only as an example to illustrate our method intended to be extended to unknown distributions, using the CLT for the mean behavior and heavy tail distributions of the Pareto type for the tail.
Since the exact distribution of the Pareto sum $S_n$ of iid Pareto rv's is known, we will be able to judge about the quality of the approximation proposed in Theorem \ref{dfSn2-cdtional} when comparing $G_{n,\a,k}$ with the exact distribution of $S_n$. We will then compare the respective associated risk measures.\\
Recall that the distribution of the sum $S_n$ of iid Pareto rv's is given by the following (see \cite{ramsay} and references therein). 
\begin{itemize}
\item[$\triangleright$] For $0<\a<2$ and $\a\neq 1$, the pdf $f_n$ of $S_n$ is given explicitly by the series expansion (see Brennan {\it et al.}, 68, and Blum, 70 \cite{blum})
\begin{equation}\label{blum}
f_n(x)=\frac{-1}{\pi}\sum_{j=1}^n \binom{n}{j}\big(-\Gamma(1-\a)\big)^j \sin(\pi\a j)\sum_{m=0}^\infty C_{n-j,m}\frac{\Gamma(m+\a j+1) }{x^{m+\a j+1}}
\end{equation}
where $C_{k,m}$ is the $m$th coefficient in the series expansion of the $k$th power of the confluent hypergeometric function: 
$\d \sum_{m=0}^\infty C_{k,m}t^m=\left(\sum_{j=0}^\infty \binom{-\a}{j-\a}\frac{t^j}{j!}\right)^k$ 
but computational difficulties may arise for large values of $n$ and certain ranges of $x$ and $\a$, as pointed out in \cite{blum}.
\item[$\triangleright$] An alternative method, based on the inversion of the Laplace transform, has been proposed in \cite{ramsay} and provides an explicit expression as well in the case $\a\in\N^*$ and for a Pareto $H_\b$ (ie not only for the case $\beta=1$). We have
\begin{equation}\label{ramsay}
f_n(t)= \frac1{n\b}\int_0^\infty \g_{m,n}(v/n)e^{-\big(1+\frac{t}{n\b}\big)v}dv
\end{equation}
where, for $v>0$, 
\begin{eqnarray*}
 \g_{m,n}(v)&:=&(-1)^{n+1}m^n \sum_{j=0}^{[(n-1)/2]} (-\pi^2)^j \binom{n}{2j+1}\big(Ei_{m+1}(v)\big)^{n-2j-1}\left(\frac{v^m}{m!}\right)^{2j+1}\\
 Ei_{m+1}(x)&:=&\frac{x^{m}}{m!}\left(\gamma+\ln x-\sum_{j=1}^{m}\frac1{j} +  \sum_{j\ge0, j\neq m}\frac{x^j}{(j-m)j!}\right)
 \end{eqnarray*}
 $\gamma$ being the Euler constant.
\item[$\triangleright$] The pdf  $f_n$ of $S_n$, satisfying $f_n=f^{n\ast}$, can also be simply evaluated numerically using the recursive convolution equation 
\begin{equation}\label{recursiveConvol}
f^{j\ast }(x)=f^{\ast (j-1)}\ast f (x)=\int_0^x f^{ (j-1)\ast}(x-u)f(u)~du, \quad \text{for} ~j\ge 2,
\end{equation}
and $f^{1\ast}=f$. This recursive approach may yield good results, but is relatively slow for large values of  $n$ and $x$. 
\end{itemize}
\item The convolution product $h_y^{ (k-1)\ast}$ appearing in $G_{n,\a,k}$ can be numerically evaluated using, either the recursive convolution equation \eqref{recursiveConvol} applied to $h$ (note that $k-1$ being small, there is no problem of convergence anymore), or, for $\a=1,2$, the explicit expression \eqref{ramsay} when replacing $\beta$ by $y$.
\item Finally recall the following result by Feller (see \cite{feller}) on the convolution closure of distributions with regularly varying tails, which applies in our Pareto example but may also be useful when extending the method.
\begin{lem} \label{feller-convol}
If $F_1$ and $F_2$ are two cdfs with regularly varying tails with tail index $\b\ge 0$, then the convolution  $F_1\ast F_2$ is also regularly varying with the same tail index $\b$.
\end{lem}
Note that this lemma implies the result given in Lemma \ref{tailMax}.\\

\n As a consequence of Lemma \ref{feller-convol} in our Pareto case, we have
\begin{equation}\label{approximTailPareto}
\int_x^{\infty} h_y^{(k-1)\ast} (u)du ~ \underset{x\to\infty}{\sim} ~ (k-1) \int_x^{\infty} h_y (u)du 
\end{equation}
\end{enumerate}

~\\
\n{\it Proof of Theorem \ref{dfSn2-cdtional}}\\

\n $\triangleright$ 
Let us express  the cdf of $S_n$. Note that $\P(S_n \le x)=\P(1\le S_n \le x)$. For any $x\ge 1$, we can write
$$
\P(S_n \le x) = \int_1^x \P\Big( T_k +U_{n-k+1} \le x -y ~/~ X_{(n-k+1)}=y \Big)f_{(n-k+1)}(y)dy 
$$
Hence, if $k=1$,
\begin{eqnarray}\label{theo-k1}
\P(S_n \le x)  
&=& \int_1^x  f_{(n)}(y)\int_0^{x-y}  f_{T_1/X_{(n)}=y } (v)dv ~dy 
\end{eqnarray}
and, for $k\ge 2$, using the conditional independence of $T_k /X_{(n-k+1)}$ and $U_{n-k+1}/X_{(n-k+1)}$, 
\begin{eqnarray}\label{theo-k2}
\P(S_n \le x)  
&=&  \int_1^x  f_{(n-k+1)}(y) \int_0^{x-y} f_{T_k /X_{(n-k+1)}=y} \ast f_{U_{n-k+1} /X_{(n-k+1)}=y}(v) dv ~dy \qquad
\end{eqnarray}

\n The next two steps consist in evaluating  the limit distribution of $T_k ~/~(X_{(n-k+1)}=y)$ and the exact distribution of $U_{n-k+1}~/~(X_{(n-k+1)}=y)$.\\

\n $\triangleright$ {\it A limiting normal distribution for $T_k /(X_{(n-k+1)}=y)$}
\begin{prop}\label{conditCLT}
The conditional distribution of the trimmed sum $T_k$ defined in \eqref{eq:decompSn} given the $(n-k+1)$th largest rv $X_{(n-k+1)}$ can be approximated, for large $n$,  by the normal distribution $\d {\cal N}\Big(m_1(\a,n,k,y), \sigma^2(\a,n,k,y)\Big)$: 
\begin{equation}\label{clt-cdtional-trimmed}
{\cal L} \Big(T_k /(X_{(n-k+1)}=y)\Big) ~\underset{n\to\infty}{\overset{d}{\sim}}  ~ {\cal N}\Big(m_1(\a,n,k,y), \sigma^2(\a,n,k,y)\Big) 
\end{equation}
with $y>1$ and where the mean  $m_1(\a,n,k,y)$ and  the variance $\sigma^2(\a,n,k,y)$ are defined in \eqref{m1condi} and \eqref{sigma2condi} respectively. \\~
\end{prop}

\n {\it Proof of Proposition \ref{conditCLT}}\\

\n Since $T_k /X_{(n-k+1)}~$  has the same distribution as $\sum_{j=1}^{n-k} Y_j$ with $(Y_j)$ an $(n-k)$-sample with parent cdf 
defined by $\d F_Y(.)=\P\Big(X_i\le .~/~ X_i< X_{(n-k+1)}\Big)$, we may apply the CLT whenever the 2nd moment of $Y_j$ is finite. Note that for the reasons explained previously, we will choose $p\ge 4$  for a better fit (after noticing that if $X_{(i)}$ has a finite $p$th moment, for $i\le n-k$, so does $X/X<X_{(n-k+1)}$).\\
We need to compute the first two moments of $T_k /(X_{(n-k+1)}=y)$, $m_1(\a,n,k,y)$ and $m_2(\a,n,k,y)$ respectively. We do it when considering the sum itself, since it involves less computations than the direct computation. Indeed, using \eqref{cdtionalMean}, \eqref{cdtionalE2} and \eqref{cdtionalEij} respectively, and applying the multinomial theorem, lead to, for $k>p/\a -1$, 
\begin{eqnarray*}
m_1(y)&:=&m_1(\a,n,k,y):=  \sum_{j=1}^{n-k}\E\big(X_{(j)}/X_{(n-k+1)}=y\big) \\
&=& (n-k)\int_0^1 \!\!\! F^{\leftarrow}\Big(uF(y)\Big)\sum_{j=0}^{n-k-1} 
\left(\!\!\!\begin{array}{c}
n-k-1\\
j
\end{array}\!\!\!\right)
 ~ u^{j}(1-u)^{n-k-1-j} ~du \\
&& \qquad \qquad \text{where } \left(\!\!\!\begin{array}{c}
n-k-1\\
j
\end{array}\!\!\!\right) ~\text{denotes the binomial coefficient}\\
&=& ~(n-k)\int_0^1 \!\!\! F^{\leftarrow}\Big(uF(y)\Big)~du \quad (\text{using the binomial theorem})
\end{eqnarray*}
{\it i.e.} when considering the $\a$-Pareto distribution, using \eqref{df-InversePareto},
\begin{eqnarray*}  
 m_1(y)&=& (n-k(\a))\int_0^1 \big(1-uF(y)\big)^{-1/\a}du  \\
&=& \frac{n-k(\a)}{F(y)} \times \left\{
 \begin{array}{lc}
 \displaystyle \frac{1-\big(\overline{F}(y)\big)^{1-1/\a}}{1-1/\a} & \text{if}~\a\neq 1\\
 &\\
 |\ln \overline F(y)~| & \text{if}~\a=1
 \end{array}
 \right.
\end{eqnarray*}
hence \eqref{m1condi}.\\
Let us compute the 2nd moment $m_2(y):=m_2(\a,n,k,y)$, introducing the notation $a=F(y)$.
\begin{eqnarray*}
m_2(y)&=& \sum_{j=1}^{n-k} \E\big( X_{(j)}^2/X_{(n-k+1)}=y \big) + 2\sum_{j=2}^{n-k} \sum_{i=1}^{j-1} \E\big(X_{(i)}X_{(j)} /X_{(n-k+1)}=y \big) \\
&=& ~(n-k)\int_0^1 \!\!\! \Big(F^{\leftarrow}(au)\Big)^2 du  ~ + ~2 (n-k)(n-k-1)\int_0^1 F^{\leftarrow}(au) \int_u^1 F^{\leftarrow}(av) \\
&& \qquad \times \sum_{j=0}^{n-k-2} \sum_{i=0}^{j}
\left(\!\!\!\begin{array}{c}
n-k-2\\
i , j-i ,n-k-2-j
\end{array}\!\!\!\right)
 ~ u^{i} \big(v-u\big)^{j-i}\big(1-v \big)^{n-k-2-j} dv~du \\
 &&\qquad \quad \text{where } \left(\!\!\!\begin{array}{c}
n-k-2\\
i , j-i ,n-k-2-j
\end{array}\!\!\!\right) ~\text{denotes the multinomial coefficient}\\
&=& ~(n-k)\left\{\int_0^1 \!\!\! \Big(F^{\leftarrow}(au)\Big)^2 du  + 2 (n-k-1)\int_0^1 \!\! F^{\leftarrow}(au) \int_u^1 F^{\leftarrow}(av) dv~du \right\} \\
&& (\text{using the multinomial theorem})
\end{eqnarray*}
Hence, for Pareto, via \eqref{df-InversePareto}, it comes 
\begin{eqnarray*}
m_2(y)&=& (n-k)\left\{\int_0^1 (1-au)^{-2/\a}du  + 2 (n-k-1) \int_0^1  (1-au)^{-1/\a}  \int_u^1 (1-av)^{-1/\a} dv~du \right\} \\
&=& ~(n-k)\int_0^1 (1-au)^{-2/\a}du ~ +~(n-k)(n-k-1)\times\\
&&\quad \left\{
\begin{array}{lc}
\displaystyle \frac2{a(1-1/\a)}  \left( \int_0^1  (1-au)^{1-2/\a}du~ - 
~(1-a)^{1-1/\a}  \int_0^1 (1-au)^{-1/\a} du \right)& \text{if}~\a\neq 1\\
&\\
\d \frac{\ln ^2(1-a)}{a^2}& \text{if}~\a=1
\end{array}
\right.  
\end{eqnarray*}
\begin{eqnarray*}
\mbox{{\it i.e.}} \quad  m_2(y) &=& \frac{n-k(\a)}{F(y)}\times \left\{
\begin{array}{lc}
\displaystyle \frac{1-\big(\overline{F}(y)\big)^{1-2/\a}}{1-2/\a} & \text{if}~\a\neq 2\\
&\\
\displaystyle |\ln \overline F(y)| & \text{if}~\a=2
\end{array}
\right. 
  \nonumber \\  
&&   \quad + ~\frac{\big(n-k(\a)\big)\big(n-k(\a)-1\big)}{F^2(y)} \times\left\{
\begin{array}{lc}
\displaystyle \frac{\big(1-(\overline{F}(y))^{1-1/\a}\big)^2}{(1-1/\a)^2} & \text{if}~\a\neq 1\\
&\\
\displaystyle \ln ^2\big(\overline F(y)\big)& \text{if}~\a=1
\end{array}
\right. \\
\end{eqnarray*}
hence \eqref{sigma2condi}.\\
Then the result given in Proposition \ref{conditCLT} follows. $\Box$\\

\n $\triangleright$ {\it A Pareto distribution  for the conditional sum of the largest order statistics}\\

\n  Now, let us look at $U_{n-k+1} /\big(X_{(n-k+1)}=y\big)$, assuming $k\ge 2$.  Its distribution may be computed explicitly via \eqref{cdtionalSuccessivepdf} that becomes, when taking $i=n-k$ and $j=k$, and for $y\le x_1\le \ldots\le x_{k-1}$,
$$
f_{X_{(n-k+2)},\ldots , X_{(n)}/X_{(n-k+1)}=y}(x_1,\ldots, x_{k-1})=\frac{(k-1)!}{\big(1-F(y)\big)^{k-1}}~\prod_{l=1}^{k-1}f(x_l)
= \frac{(k-1)!~\a^{k-1}}{y^{-\a(k-1)}}~\prod_{l=1}^{k-1}x_l^{-\a-1}
 $$
 \begin{eqnarray*}
 \text{\it i.e.} \quad f_{X_{(n-k+2)},\ldots , X_{(n)}/X_{(n-k+1)}=y}(x_1,\ldots, x_{k-1})= (k-1)!~\prod_{l=1}^{k-1}h_y(x_l)\1_{(x_1\le \ldots\le x_{k-1})}
 \end{eqnarray*}
 where $h_y$ is the probability density function (df) a Pareto rv with parameters  $\a$ and $y$.\\
We can then deduce, taking into account the number of possible permutations, that the conditional density of the sum $U_{n-k+1}$ given that  $(X_{(n-k+1)}=y)$ is defined, for any $s\ge (k-1)y$, by  
\begin{equation}\label{cdfUcdtional}
f_{U_{n-k+1} /(X_{(n-k+1)}=y)}(s)=  h_y^{ (k-1)\ast} (s) 
\end{equation}
where $ (k-1)\ast$ denotes the convolution product of order $k-1$.\\
Note that we could have retrieved this conditional density, noticing, as previously for $T_k$, that $U_{n-k+1} / X_{(n-k+1)}$ can be written as 
$\d 
U_{n-k+1} / X_{(n-k+1)} \overset{d}{=} \sum_{j=1}^{k-1} Z_j 
$
where the $Z_j$ are i.i.d. rv's  with parent rv Z and parent cdf defined by 
$\d F_Z(.)=\P\Big[X\le .~/~ X> X_{(n-k+1)}\Big]$.\\
 
\n $\triangleright$ Combining Proposition \ref{conditCLT}, the result \eqref{cdfUcdtional}, and \eqref{theo-k1}, \eqref{theo-k2}, allows to conclude to Theorem \ref{dfSn2-cdtional}. \hfill $\Box$

\subsection{On the quality of the approximation of the distribution of the Pareto sum $S_n$} \label{rateCv} 

\n To estimate the quality of the approximation of the distribution of the Pareto sum $S_n$, we compare analytically the exact distribution of $S_n$ with the distribution $G_{n,\alpha;k}$ defined in Theorem \ref{dfSn2-cdtional}. 
It could also be done numerically, as for instance in \cite{furrer} with the distance between two distributions $F$ and $G$ defined by $ d_i(F,G)=\int_{1}^{\infty}\big| F(x)-G(x) \big |^i dx $, with $i=1,2$. We will proceed numerically only when considering the tail of the distributions, estimating the distance in the tails through the VaR measure (see \S \ref{numerical}).
When looking at the entire distributions, we will focus on the analytical comparison mainly for the case $\a>2$ (with some hints for the case $\a\le 2$). 
Note that it is not possible to compare directly the expressions of the VaR corresponding to, respectively, the exact and approximative distributions, since they can only be expressed as the inverse function of a cdf. Nevertheless, we can compare the tails of these two distributions to calibrate the accuracy of the approximative VaR since
$$
\left| \P(S_n >  x) - \overline G_{n,\alpha;k}(x)\right |=\left| \P(S_n \le x) - G_{n,\alpha;k}(x)\right |
$$

\n Moreover, we will compare analytically our result with a normal approximation made on the entire sum (and not the trimmed one) since, for $\a>2$, the CLT applies and, as already noticed, is often used in practice. \\

\n 
Since Normex uses the exact distribution of the last upper order statistics, comparing the true distribution of $S_n$ with its approximation $G_{n,\a;k}$ simply comes back to the comparison of the true distribution of $n-k$ iid rv's with the normal distribution (when applying the CLT).  
Note that, when extending Normex to any distribution, an error term should be added to this latter evaluation. It comes from the approximation of the extremes distribution by a Pareto one.\\ 


\n Suppose $\alpha > 2$. 
\n Applying the CLT gives the normal approximation $\d {\cal N}(\mu_n, s_n^2)$, with $\mu_n:=\E(S_n)$ and  $\d s_n^2:=var(S_n)$, where in the case of a Pareto sum, $\mu_n=\frac{n\a}{\a-1}$, and $\d s_n^2= \frac{n\a}{(\a-1)^2(\a-2)}$. 
We know that applying the CLT directly to $S_n$ leads to non satisfactory results, not only for the estimation of risk measures but even for the mean behavior, since for any $x$,  the quantity $Q_1(x)$, involving the 3rd moment of $X$, appearing in the error \eqref{edgeworth} made when approximating the exact distribution of $S_n$ by a normal one, is infinite for any $2<\a\le 3$. The rate of convergence in $n$ is reduced to $O(1)$. 
When $\a>3$, even if  the rate of convergence improves because $Q_1(x)<\infty$, we still have $Q_2(x)=\infty$ (because the 4th moment of $X$ does not exist), which means that we cannot get a rate of order $1/n$.\\
Now let us look at the rate of convergence when approximating $S_n$ with $G_{n,\alpha;k}$. \\ 

\n Recall \eqref{theo-k1} and \eqref{theo-k2}.
Considering the exact distribution of the Pareto sum $S_n$ means taking,  at given $y>1$ and for any $k\ge 1$, 
$ T_k \le x -y ~/~( X_{(n-k+1)}=y) ~ \overset{d}{=} ~\sum_{j=1}^{n-k}Y_{j} $ with $Y_{j}$ iid rv's with parent rv $Y$ with finite $p$th moment and pdf $g$ defined, for $z\le (n-k)y$, by
\begin{equation}\label{g}
f_{T_k/X_{(n-k+1)}=y}(z) = g^{(n-k)\ast}(z)  \quad\mbox{with}~g(u)= \frac{\a}{F(y)} ~ u^{-\a-1}  \1_{(1\le u \le y)}
\end{equation}
Considering our approximation means to replace  $f_{T_k/X_{(n-k+1)}=y}$ by the pdf $\varphi_{m_{1k}, \sigma^2_k}$, of the normal distribution $\d {\cal N}\Big(m_1(\a,n,k,y), \sigma^2(\a,n,k,y)\Big)$ defined in Proposition \ref{conditCLT}.\\

\n Let us look at the three first moments of $Y$. 
Although the direct dependence is on $\a$ (and $y$) and only indirectly on $k$ since $k=k(\a)$, we introduce $k$ in the index notation for convenience and have
\begin{equation}\label{muy}
\mu_{y}:=\E(Y)=\frac{m_{1}(\a,n,k;y)}{n-k}
= \frac1{1-1/\a}\times \frac{1-y^{1-\a}}{1-y^{-\a}}\1_{(\a\neq 1,2)}+\frac{\ln(y)}{1-y^{-1}}\1_{(\a=1)}+\frac2{1+y^{-1}}\1_{(\a=2)}
\end{equation}
(note that $\mu(y)>1$, for any $\a$ that we consider, and any $y>1$), and 
\begin{eqnarray}\label{gy}
\gamma^2_{y}:=var(Y)= \frac{ \sigma^2(\a,n,k;y)}{n-k} \!\!\! &=&\!\!\! 
\frac1{1-y^{-\a}} \left(\frac{1-y^{2-\a}}{1-2/\a} -\frac1{(1-1/\a)^2}\times\frac{(1-y^{1-\a})^2}{1-y^{-\a}} \right)\1_{(\a\neq 1,2)}\quad \\
\!\!\!&&\!\!\! + ~ y \left( 1-\frac{y\ln^2 (y)}{(y-1)^2}\right) \1_{(\a=1)} ~ +~ 2 ~\frac{y^2}{y^2-1}\left( \ln(y)-2~\frac{y-1}{y+1}\right) \1_{(\a=2)}  \nonumber
\end{eqnarray}
We also need to compute the third moment. A straightforward computation provides
$$
\E(|Y- \mu_{y}|^3)= \frac{\a}{1-y^{-\a}}\left[2h(\mu)-h(1)-h(y) \right]
$$
where $h$ denotes the antiderivative of the function $H(z)= (\mu^3-3\mu^2 z+3\mu z^2-z^3)z^{-\a-1}$,  \\
{\it i.e.},~if $\a\neq 1,2$,
\begin{eqnarray*}
\E(|Y- \mu_{y}|^3) &=&  
 \frac{\a}{1-y^{-\a}}\left[\frac{\1_{(\a\neq 3)}}{3-\a}y^{3-\a} +\1_{(\a= 3)}\ln(y)+\frac{3 \mu_{y}}{\a-2}y^{2-\a}-\frac{3 \mu^2_{y}}{\a-1}y^{1-\a}+\frac{\mu^3_{y}}{\a}y^{-\a} + \right.\\
\!\! &\!\! &\!\! \!\! \left.\frac{12 ~\mu^{3-\a}_{y}~\1_{(\a\neq 3)}}{\a(\a-1)(\a-2)(\a-3)} -\!\! \left(2\ln  \mu_{y} +\frac{11}{3}\right)\! \1_{(\a= 3)}
+\frac{ \mu^3_{y}}{\a}-\frac{ 3\mu^2_{y}}{\a-1}+\frac{3 \mu_{y}}{\a-2}-\frac{\1_{(\a\neq 3)}}{\a-3} \right]
\end{eqnarray*}
whereas, if $\a=1$,
\begin{eqnarray*}
\E(|Y- \mu_{y}|^3) &=& \frac1{(1-y^{-1})^2} \left[ 
\frac{y^2}{2} - y~\left(\frac12 + 3\ln(y)\right) +\frac{\ln^3(y)}{1-y^{-1}}\left(3+ \frac1{y-1}+\frac1{1-y^{-1}}\right) 
\right.\\
&& \left.\qquad\quad + ~ \frac{3\ln^2(y)}{1-y^{-1}}\left[1-2\ln_2(y)+2\ln(1-y^{-1}) \right]-3\ln(y)+\frac{1-y^{-1}}{2}
\right] 
\end{eqnarray*}
and, if $\a=2$,
\begin{eqnarray*}
\E(|Y- \mu_{y}|^3) &=&  \frac4{1-y^{-1}} \left[ \frac{2y}{1+y^{-1}} -3\ln(y)-\frac6{1+y}+\frac2{(1+y)^2}+\frac{1+y^{-1}}{2} \right.\\
&&\left.  \qquad\quad + ~ 3\left(1+2\ln2-2\ln(1+y^{-1})\right) - \frac6{1+y^{-1}}+\frac2{(1+y^{-1})^2} \right]\\
&&
\end{eqnarray*}

\n For simplicity, let us look at the case $2<\a\le 3$ 
and consider the Berry-Ess\'een  inequality.  For $\a>3$, we would use the Edgeworth expansion, with similar arguments as developed below.  
Various authors have worked on this type of Berry-Ess\'een inequality, in particular to sharpen the accuracy of the constant appearing in it. In the case of  
 Berry-Ess\'een  bounds, the value of the constant factor $c$ has decreased from 7.59 by Ess\'een  (1942)  to 0.4785 by Tyurin (2010; \cite{tyurin}), to 0.4693 by Shevtsova (2012, \cite{shev}; see also \cite{ko:she} for a detailed review), in the iid case, and to 0.5600 in the general case. 
Note also that these past decades, much literature has been dedicated to the generalization of this type of inequality; we will not provide exhaustive references, besides pointing out the remarkable contribution by Stein (1972 \cite{stein}, 1986 \cite{stein86}) who proposed a uniform upper bound to the normal approximation as in the Berry-Ess\'een  bound, but under general distributional assumptions, allowing dependent and nonidentical distributions; the Stein method has been used to develop many studies, in particular by Chen \& Shao (2004, \cite{chen:shao}) to obtain sharp bounds of the Berry-Ess\'een  type under local dependence (see also \cite{pinelis} for new developments).\\~\\
Since $\a>2$, we only have to consider the case $k=1$ (see Table 2).
We can write
$$
|\P(S_n \le x) - G_{n,\alpha;1}(x) | ~ \le ~   \int_1^x |\P\Big( T_1 \le x -y ~/~ X_{(n)}=y \Big) - \Phi_{(n-1)\mu_{y}, (n-1)\g^2_{y}} (x-y) | f_{(n)}(y)dy
$$
Since the conditions on moments of $Y$ are satisfied, we can use the Berry-Ess\'een  inequality to provide a non-uniform bound of the error made when approximating the exact distribution by $ G_{n,\alpha;1} $. 
Indeed, we have
\begin{eqnarray} \label{nonun1bis}
&& \left| \P\Big( T_1 \le x -y ~/~ X_{(n)}=y \Big) - \Phi_{(n-1)\mu_{y}, (n-1)\g^2_{y}}(x-y) \right | \nonumber \\
&& = ~  \left |  \P\left(\frac{\sum_{i=1}^{n-1}Y_{i} -(n-1)\mu_{y} }{\sqrt{n-1}~\gamma_{y}}\le  \frac{x-y -(n-1)\mu_{y} }{\sqrt{n-1}~\gamma_{y}}\right) -
 \Phi\left(\frac{x-y -(n-1)\mu_{y} }{\sqrt{n-1}~\gamma_{y}}\right)  \right |  \nonumber \\
&& \le ~  \frac{c~C(y)}{\sqrt{n-1}} \times \frac1{\left(1+\left | \frac{x-y -(n-1)\mu_{y} }{\sqrt{n-1}~\gamma_{y}} \right | \right)^3}  
\end{eqnarray}
where $\mu_y$ and $\gamma_y$ are defined in \eqref{muy} and \eqref{gy} respectively, and 
\begin{equation} \label{Cy}
C(y):=\frac{\E(|Y- \mu_{y}|^3)}{\gamma_{y}^{3}} , \quad\text{with} ~ \E(|Y- \mu_{y}|^3) ~ \text{computed above.}
\end{equation}
We can deduce that, for any $x\ge 1$,  
\begin{eqnarray} \label{bdAnalytK}
| \P(S_n \le x) - G_{n,\alpha;1}(x) | & \le & K(x):=\frac{c}{\sqrt{n-1}} \int_1^x  \!\!\! \frac{C(y)}{\left(1+\left | \frac{x-y -(n-1)\mu_{y} }{\sqrt{n-1}~\gamma_{y}} \right | \right)^3}~ f_{(n)}(y) ~dy \qquad
\end{eqnarray}
We can compute numerically the function $K$ in \eqref{bdAnalytK}, as there is no known analytical solution for the antiderivative. We use the software R for that.\\
We  represent the bound $K$ as au function of $x$ for various $\a\in(2;3]$ and various $n$ in the figures below.\\
\begin{minipage}{0.30\linewidth}
\centering
\includegraphics[width=5cm,height=5cm]{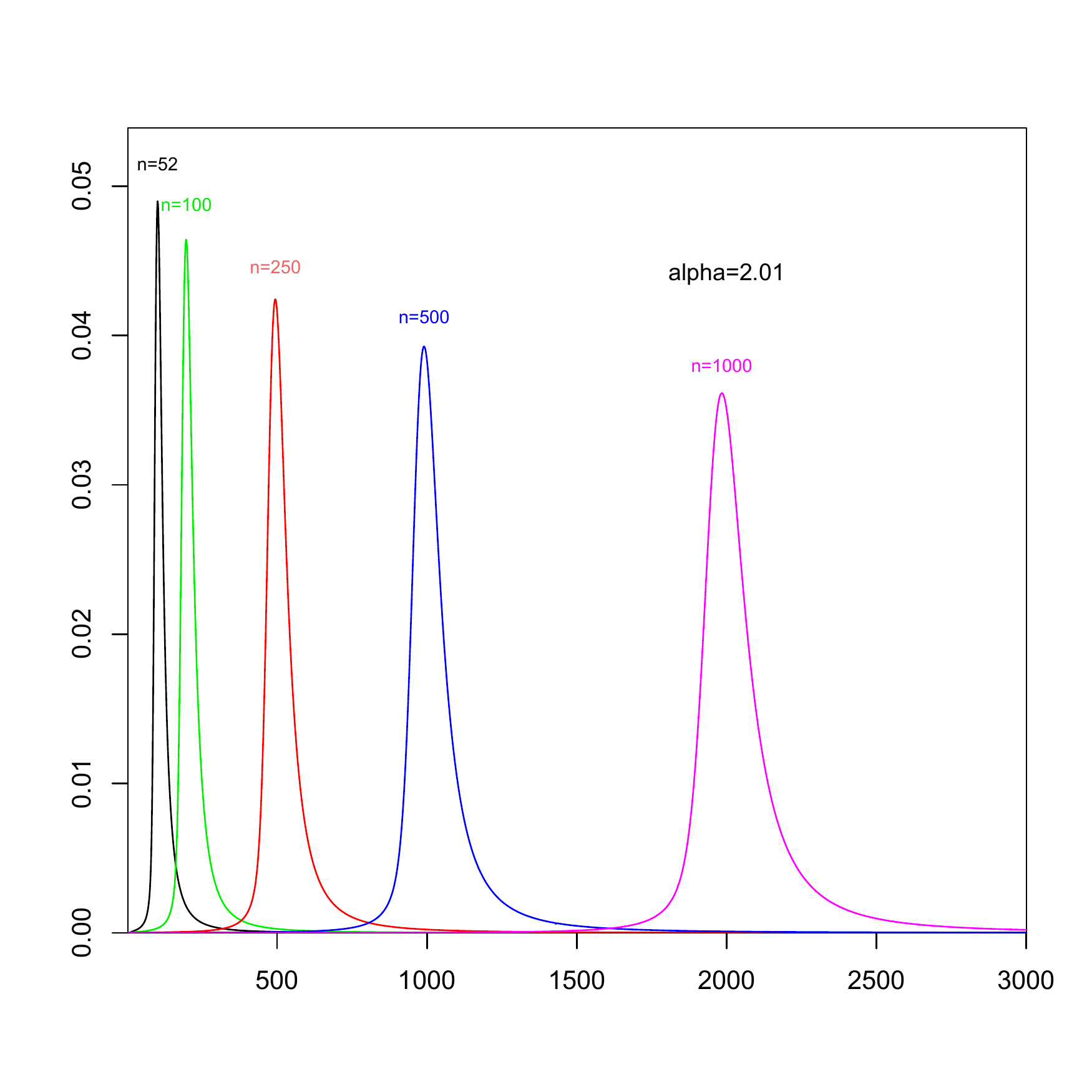}\end{minipage}
\hfill
\begin{minipage}{0.30\linewidth}
\centering
\includegraphics[width=5cm,height=5cm]{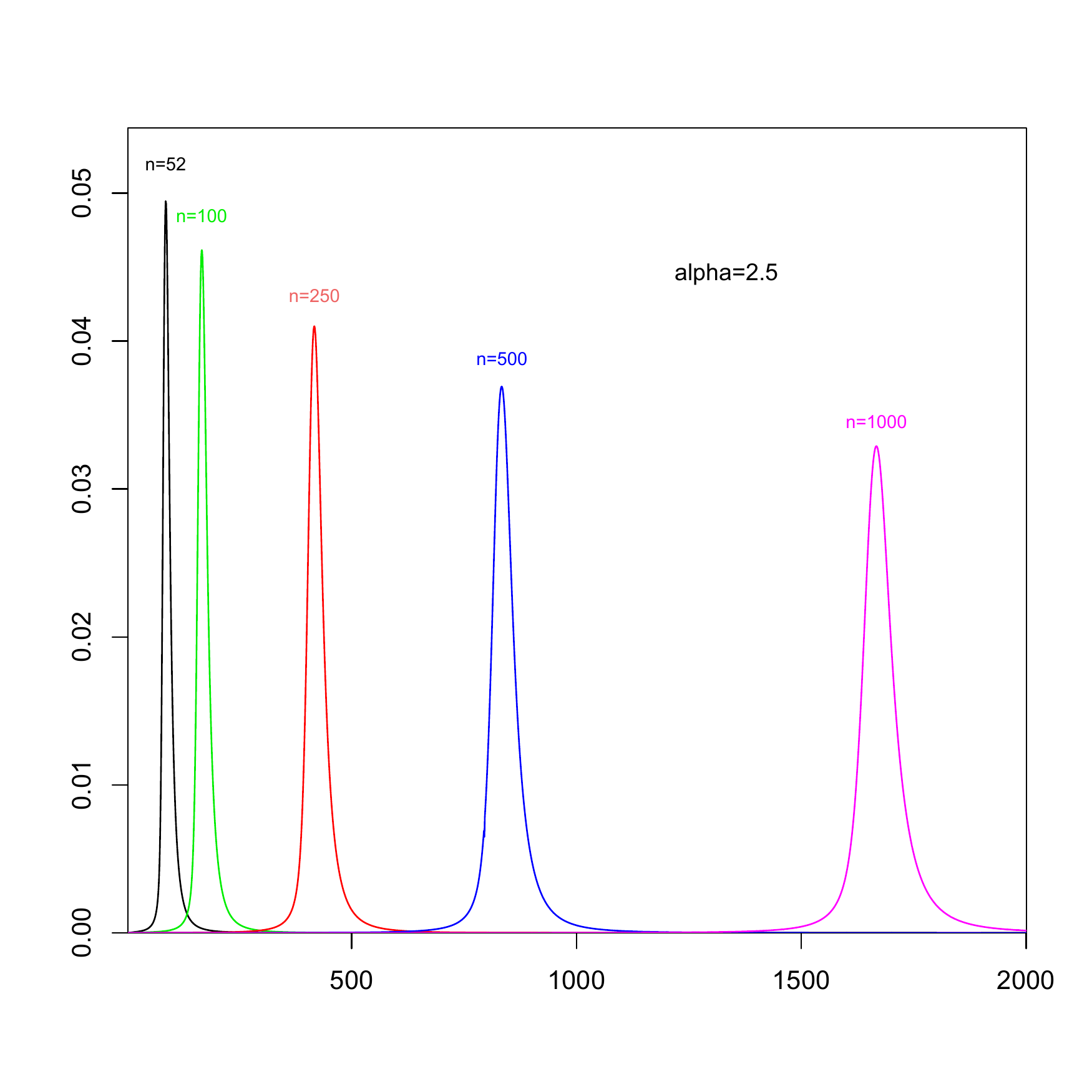} 
\end{minipage}
\hfill
\begin{minipage}{0.30\linewidth}
\centering
\includegraphics[width=5cm,height=5cm]{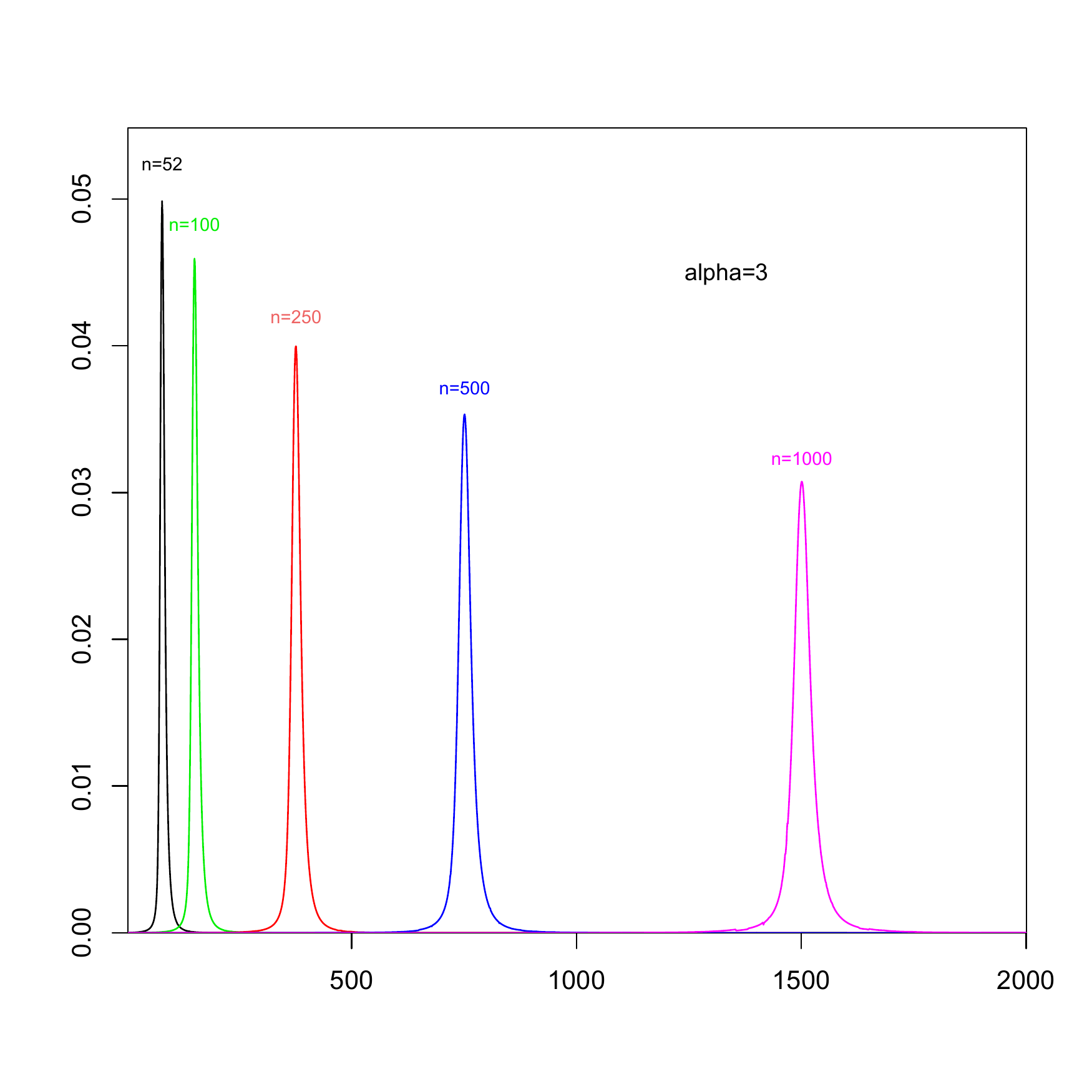} 
\end{minipage}
 \vspace{-2ex}
 \begin{center}
{\small {\it Graph of the function $K$ for various $\a$ and $n$}}
\end{center}
 
\n The bound $K$ is an increasing then decreasing function of $x$,  with a maximum less than $5\%$, which is decreasing with $n$ and $\a$. The $x$-coordinate of the maximum is proportional to $n$, the proportion decreasing with $\a$. 
The interval on the $x$-axis for which the error is larger than $1\%$ has a small amplitude, which is decreasing with $\a$.\\
We show in Table~\ref{tbl-max} the values of the coordinates $(x_{\text{max}}, y_{\text{max}})$ of the maximum of $K$ computed on R for $\a=2.01,2.5,3$ and $n=52,250,500$ respectively. \\
\begin{table}[h]
\begin{center}
\begin{tabular}{r||c|c||c|c||c|c}
   & \multicolumn{2}{c||}{$\a=2.01$}&\multicolumn{2}{c||}{$\a=2.5$}&\multicolumn{2}{c}{$\a=3.0$} \\
  n~    &  ~~$x_{\text{max}}$~~  &  ~~$y_{\text{max}}$~~  &  ~~$x_{\text{max}}$~~  &  $ ~~y_{\text{max}}$~~  &  ~~$x_{\text{max}}$~~  &  ~~$y_{\text{max}}$~   \\
\hline\hline 
   52~    & 101   &  4.9 \% &      86  &  4.9 \% &   78  &  4.9 \% \\
\hline
 100~    & 196   &  4.6 \% &    166  &  4.6 \% & 150  &  4.6 \%\\
\hline
  250~   & 494   &   4.2 \% &   417  &  4.1 \% & 376  &  4.0 \%  \\
\hline
 500~    &  990   &  3.9 \% &   834  &  3.7 \% & 751   &   3.5 \%\\
\hline
1000~   & 1984  &  3.6 \% & 1667 &  3.3 \% & 1501 &  3.0 \%\\
\end{tabular}
\caption{\label{tbl-max} \sf Coordinates of the maximum of $K$ as a function of $n$ and $\a$}
\end{center}
\end{table}

\n We can then conclude to the following proposition.
\begin{prop}\label{analComp}
The error between the true distribution of $S_n$ and its approximation $G_{n,\alpha;1}$ is bounded by:
\begin{equation*} 
| \P(S_n \le x) - G_{n,\alpha;1}(x) | \le  K(x)= \frac{c}{\sqrt{n-1}} \int_1^x  \frac{C(y)}{\left(1+\left | \frac{x-y -(n-1)\mu_{y} }{\sqrt{n-1}~\gamma_{y}} \right | \right)^3}~ f_{(n)}(y) ~dy 
\end{equation*}
with $c=0.4693$, $f_{(n)}$ computed in \eqref{pdf1OrderStat}, $C(y)$ defined in \eqref{Cy}, and $\mu_y$ and $\gamma_y$ in \eqref{muy} and \eqref{gy} respectively. \\
Moreover, for any $n\le 52$ and $\a \in (2; 3]$,  $\d 0 \le \max_{x>1} K(x) < 5\%$ and $K$ decreases very fast to 0 after having reached its maximum; the larger $n$, the faster to 0.
\end{prop}

\vspace{3ex}
\n {\bf Remark.} Let us briefly look at the case $\a \le 2$ (and $\a>1/2$).\\
 We have seen that such a case implies $k\ge 2$ (see Table 2).
 We have
\begin{eqnarray*}
\left| \P(S_n \le x) - G_{n,\alpha;k}(x) \right | & \le & \int_1^x  f_{(n-k+1)}(y)\int_0^{x-y} f_{U_{n-k+1} /X_{(n-k+1)}=y}~ \ast \\
&& \qquad\qquad\qquad \qquad \qquad \left| g^{ (n-k)\ast}  - \varphi_{(n-k)\mu_{y}, (n-k)\g^2_{y}} \right| (v) dv ~dy 
\end{eqnarray*}
Note that the Berry-Ess\'een  inequality has been proved by Petrov to hold also for probability density functions (see \cite{petrov56}, or \cite{petrov}). It has been refined by Shevtsova (see \cite{shev07}) and we will use her result to evaluate $\left| g^{ (n-k)\ast}  - \varphi_{(n-k)\mu_{y}, (n-k)\g^2_{y}} \right| $. 
So we need to go back to the pdf of the standardized sum $\d \sum_{i=1}^{n-k} \frac{Y_{i} -\mu_{y} }{\sqrt{n-k}~\gamma_{y}}$ of iid rv's with pdf $\tilde g$, which can be expressed as 
$$
\tilde g^{(n-k)\ast} \quad \text{where} \quad g(.)=\frac1{\sqrt{n-k}~\gamma_{y}} \tilde g \left(\frac{.~ -\mu_{y}}{\sqrt{n-k}~\gamma_{y}} \right)
$$
It is straightforward to show by induction that 
$$
g^ {(n-k)\ast}(v) =\frac1{\sqrt{n-k}~\gamma_{y}} \tilde g^{ (n-k)\ast}\left(\frac{v - (n-k)\mu_{y}}{\sqrt{n-k}~\gamma_{y}} \right)
$$
Then, since $\d \varphi_{a,b^2}(x)=\frac1{b} \varphi\left(\frac{x-a}{b}\right)$, we can write
\begin{equation}\label{diffStd}
g^{(n-k)\ast}(v) - \varphi_{(n-k)\mu_{y}, (n-k)\g^2_{y}}(v) ~ = ~ \frac1{\sqrt{n-k}~\g_{y}} \left( \tilde g^{(n-k)\ast} - \varphi \right) 
\left(\frac{v-(n-k)\mu_{y}}{\sqrt{n-k}~\g_{y}}\right)
\end{equation}
Since we consider a sum of $(n-k)$ iid rv's $Y_i$ ($i=1,\ldots,n-k$) with parent rv $Y$ having a finite $p$th moment, we obtain via Petrov (\cite{petrov56}) and Shevtsova (\cite{shev07}) that there exists a constant $c= 0.4014$  
 such that 
\begin{equation}\label{supdiffStd}
\hspace{-0.5ex} \sup_v \left| \left( \tilde g^{(n-k)\ast} - \varphi \right) \left(\frac{v-(n-k)\mu_{y}}{\sqrt{n-k}~\g_{y}}\right)\right|=  
\sup_v \left|  \tilde g^{(n-k)\ast} (v) - \varphi (v) \right| \le \frac{c~C(y)}{\sqrt{n-k}} ~
\end{equation}
where $\d C(y)$ is defined in \eqref{Cy}.\\  
Hence, combining \eqref{diffStd} and \eqref{supdiffStd} gives
$$
\sup_v \left|  g^ {(n-k)\ast}(v)  - \varphi_{(n-k)\mu_{y}, (n-k)\g^2_{y}} (v) \right| \le \frac{c ~ C(y)}{(n-k)\g_{y}}= \frac{c}{n-k}\times \frac{\E(|Y- \mu_{y}|^3)}{ var^2(Y)}
$$
and so
\begin{eqnarray*}
\left| \P(S_n \le x) - G_{n,\alpha;k}\right | & \le & \frac{c}{(n-k)}\int_1^x \frac{C(y)}{\g_{y}} ~f_{(n-k+1)}(y)\!\! \int_0^{x-y} \!\!\!\! \int \!\!\! f_{U_{n-k+1} /X_{(n-k+1)}=y}(t)dt~ dv ~dy 
\end{eqnarray*}
with $\displaystyle f_{(n-k+1)}(y)= \frac{\a ~ n!}{(n-k)!(k-1)!} (1-y^{-\a})^{n-k}y^{-\a k-1}$ (see \eqref{pdf1OrderStat}) and
$f_{U_{n-k+1} /X_{(n-k+1)}=y}(t)=  h_y^{ (k-1)\ast} (t)$ defined in Theorem~\ref{dfSn2-cdtional}.
As in the case $k=1$, this bound can be computed numerically.  


\section{Application to risk measures and comparison}

 \subsection{Introduction}
 
\n Let us introduce the risk measures used in solvency calculations, namely the Value-at-Risk, denoted VaR, and the Expected Shortfall 
Expected Shorfall (named also Tail-Value-at-Risk) $ES$ (or TVaR), of a rv $X$ with continuous cdf $F_X$ (and inverse function denoted by 
$F_X^{\leftarrow}$).

\begin{itemize}
\item {\it Definitions.}
\begin{itemize}
\item The Value-at-Risk of order $q$ of $X$ is simply the quantile of $F_X$ of order $q$, $q\in(0,1)$:
$$ 
 VaR_q(X)=\inf\{y\in\R: P[X>y]\le 1-q\}=F_X^{\leftarrow}(q)
 $$ 
\item If  $\E|X|<\infty$, the expected shortfall ($ES$), at confidence level $q\in (0,1)$ is defined as 
$$ ES_q(X)=\frac1{1-q}\int_q^1 VaR_{\beta}(X)~d\beta \quad\text{or}\quad ES_q(X)=\E[X ~|~ X\ge VaR_q]$$ 
It can also be thought as an average over all risks exceeding $VaR_q(X)$.\\
This risk measure does depend only on the tail cdf of $X$ and satisfies \\$\d ES_q(X) \ge VaR_q(X)$.
\end{itemize}
Note that we will simplify the notation of those risk measures writing $VaR_q$ or $ES_q$ when no confusion is possible. 
\item {\it Aggregated risks}\\
When looking at aggregated risks $\sum_{i=1}^n X_i$, it is well known that the risk measure $ES$ is coherent (see \cite{arztner}).  In particular it is subadditive, {\it i.e.}
$$
ES_q\big(\sum_{i=1}^n X_i \big)~\le ~\sum_{i=1}^n ES_q(X_i)
$$
whereas VaR is not a coherent measure, because it is not subadditive. Many examples (see e.g. \cite{em:lw}, \cite{da:jssv}) can be given where VaR is superadditive, {\it i.e.}\\
$\d VaR_q\big(\sum_{i=1}^n X_i \big)~\ge ~\sum_{i=1}^n VaR_q(X_i)$. \\
We have the following property.
\begin{prop}\label{asymptSubadd}(\cite{em:lw}) \\
Consider i.i.d. rv's $X_i$, $i=1,\ldots,n$ with parent rv $X$ and cdf $F_X$. Assume they are regularly varying with tail index $\b>0$, which means that the right tail $1- F_X$ of its distribution satisfies
$$
\lim_{x\to\infty} \frac{1- F_X(ax)}{1- F_X(x)} =a^{-\b}, ~\forall a>0
$$
Then the risk measure VaR  is asymptotically subadditive for $X_1,\ldots,X_n$ if and only if $\b\ge 1$:
$$
\lim_{q\nearrow 1} \frac{VaR_q\big(\sum_{i=1}^n X_i \big)}{\sum_{i=1}^n VaR_q(X_i)} ~\le 1   \quad \Leftrightarrow \quad \b\ge 1
$$
~
\end{prop}
In particular, for large $q$, $VaR_q$ being interpreted as the risk capital,  the diversification benefit breaks down whenever $0<\b<1$.\\
~

\item {\it Pareto risks}\\
\n In the case of a $\a$-Pareto distribution, we deduce from \eqref{df-InversePareto} analytical expressions of those two risk measures, namely
\begin{equation}\label{VaR-Pareto}
VaR_q(X)=  F^{\leftarrow}(q)= (1-q)^{-\frac1\a}
\end{equation}
and, if $X\in L^1$, {\it i.e.} if $\a>1$, then
\begin{equation}\label{ES-Pareto}
ES_q(X) ~=~ \frac{\a}{(\a-1)(1-q)}(VaR_q(X))^{1-\a}= ~ \frac{\a}{(\a-1)}(1-q)^{-1/\a}
\end{equation}

\n Proposition \ref{asymptSubadd} applies when considering $\a$-Pareto iid rv's, and the risk measure VaR is asymptotically superadditive (respectively subadditive), if $\a\in(0,1)$ ($\a\ge 1$ respectively). 

\end{itemize}

\subsection{Possible approximations of VaR}

\n As an example, we treat the case of one of the two main risk measures, and choose the VaR, since it is the main one used for solvency requirement.
We would proceed in the same way for the Expected Shortfall.  \\ 

\n We deduce the approximations $z_q^{(i)}$ of the VaR of order $q$ from the various limit theorems, $(i)$ indicating the chosen method, namely $(1)$ for the GCLT approach, $(2)$ for the CLT one, $(3)$ for the max one,  $(4)$ for the Zaliapin et al.'s method, and $(5)$ for Normex. We obtain:\\

\n $\triangleright$ via the GCLT , for  $0< \a \le 2$: 
 \begin{eqnarray*} 
&&\text{- for}~\alpha<2 :\\
&& z_q^{(1)} ~ = ~  n^{1/\a}C_\a~ G^{\leftarrow}_\a(q) + b_n \quad {\small (\text{with}~ G_\a ~(\a, 1,1,0)\text{-stable distribution}) } \quad\\
&& {\small \text{for}~ 1/2<\a<2, ~ \text{and for}~ q>0.95,} \\ 
&& z_q^{(1bis)} ~ = ~ n^{1/\a}q^ {-1/\a} +b_n\\
&&\text{- for}~\alpha=2 : \\
&& z_q^{(1)} ~ = ~ d_n ~ \Phi^{\leftarrow}(q) + 2n \qquad \qquad \qquad\qquad\qquad \qquad \qquad\qquad\qquad \qquad \qquad\qquad\qquad\\
&&
\end{eqnarray*} 

\n $\triangleright$  via the CLT, for  $\a >2$:
\begin{eqnarray*}
&& z_q^{(2)} ~ = ~ \frac {\sqrt{n\a}}{(\a-1)\sqrt{\a-2}} ~ \Phi^{\leftarrow}(q) + \frac{n\a}{\a-1} \qquad\qquad\qquad \qquad\qquad\qquad \qquad\qquad \qquad \quad \\
&&
\end{eqnarray*} 

\n $\triangleright$ via the Max (EVT) approach, for high order $q$, for any positive $\a$:
\begin{eqnarray*}
&& z_q^{(3)} ~ = ~  n^{1/\a}\Big(\log(1/q)\Big)^ {-1/\a} +b_n  \qquad\qquad\qquad \qquad\qquad\qquad \qquad\qquad \qquad\qquad \qquad \\
&&
\end{eqnarray*} 

\n $\triangleright$ via the Zaliapin {\it et al.}'s method (\cite{zaliapin}), for $2/3<\a<2$:
 \begin{eqnarray*}
&& z_q^{(4)} ~ = ~  \Big( \sigma(\a,n,2)~ \Phi^{\leftarrow}(q) + m_1(\a,n,2)\Big)~+~ T^{\leftarrow}_{\a,n}(q)  \qquad\qquad\qquad \qquad\qquad\qquad \qquad\qquad \quad \\
&&\qquad\qquad\qquad  \qquad \quad {\small \text{with $T_{\a,n}$ the cdf of} ~ \big(X_{(n-1)}+X_{(n)}\big)}\\
&&
\end{eqnarray*} 

\n $\triangleright$ via Normex, for any positive $\a$:
\begin{eqnarray*}
&& z_q^{(5)} ~ = ~  G_{n,\a,k}^{\leftarrow}(q) \quad \text{with}\\
&&  {\small G_{n,\a,k}(x) = \int_1^x  \frac{f_{(n-k+1)}(y)}{\sigma(y)}  \int_0^{x-y}\!\!\! \left(\int_0^v\!\!\!  \varphi\Big(\frac{v-u-m_1(y)}{\sigma(y)}\Big) h_y^{\star (k-1)} (u)du\right)~dv ~dy }  \qquad\qquad\\
&&
\end{eqnarray*}


\subsection{Numerical study - comparison of the methods}\label{numerical}

Since there is no explicit analytical formula for the true quantiles of $S_n$, we will complete the analytical comparison of the distributions of $S_n$ and $G_{n,\a,k}$ given in $\S$\ref{rateCv},  providing here a numerical comparison between the quantile of $S_n$ and the quantiles obtained by the various methods seen so far.

\subsubsection{Presentation of the study}

\n We simulate  $(X_i, i=1,\ldots,n)$ with parent r.v. $X$ $\a$-Pareto distributed, with different sample sizes, varying from $n=52$ (corresponding to aggregating weekly returns to obtain yearly returns), through $n=250$  (corresponding to aggregating daily returns to obtain yearly returns),  to $n=500$ representing a large size portfolio. \\

\n We consider different shape parameters, namely $ \a= 3/2 ;  2 ;  5/2 ;  3 ;  4 $, respectively.\\
Recall that simulated Pareto rv's $X_i$'s ($i\ge 1)$ can be obtained simulating a uniform rv $U$ on $(0,1]$ then applying the transformation $X_i=U^{-1/\a}$.  \\

\n For each $n$ and each $\a$, we aggregate the realizations $x_i$'s ($i=1,\ldots,n$). We repeat the operation $N=10^7$ times, thus obtaining $10^7$ realizations of the Pareto sum $S_n$, from which we can estimate its quantiles. \\

\n Let $z_q$ denotes the empirical quantile of order $q$ of the Pareto sum $S_n$ (associated with the empirical cdf $F_{S_n}$ and pdf $f_{S_n}$), defined by
$$
z_q=:=\inf\{t~|~F_{S_n}(t)\ge q\}, \quad\mbox{with}~0<q<1.
$$
Recall, for completeness, that the empirical quantile of $S_n$ converges to the true quantile as $N\to\infty$ and has an asymptotic normal behavior, from which we deduce the following confidence interval at probability a for the true quantile:
\begin{equation}\label{bounds-CI-simul-zq}
z_q \pm \Phi^{\leftarrow}(a/2)\times\frac{\sqrt{q(1-q)}}{f_{S_n}(q)\sqrt{N}} \qquad
\end{equation}
where $f_{S_n}$ can be empirically estimated for such a large $N$. We do not compute them numerically:  $N$  being very large, bounds are close.\\

\n We compute the values of the quantiles  of order $q$, $z_q^{(i)}$ ($(i)$ indicating the chosen method), obtained by the three main methods, the GCLT method, the Max one, and Normex, respectively. We do it for various values of $\a$ and $n$. We compare them with the (empirical) quantile $z_q$ obtained via Pareto simulations (representing the true quantile). For that, we introduce the approximative relative error:
$$
\delta^{(i)}=\delta^{(i)}(q)=\frac{z_q^{(i)}} {z_q} -1
$$
We consider three possible order $q$: $95\%$, $99\%$ (threshold for Basel II) and $99.5\%$  (threshold for Solvency 2).\\

\n We use the software R to perform this numerical study, with different available packages (see the appendix).  Let us particularly mention the use of  the procedure {\it Vegas} in the package {\it R2Cuba} for the computation of the double integrals. This procedure turns out not to be always very stable for the most extreme quantiles, mainly for low values of $\a$. In practice, for the computation of integrals, we would advise to test various procedures in R2Cuba (Suave, Divonne and Cuhre, besides Vegas) or to look for other packages. Another possibility would be implementing the algorithm using  all together a different software, as e.g. Python.   


\subsubsection{Estimation of the VaR with the various methods}\label{tablesVaR}

All results obtained for various $n$ and $\a$ are given in \cite{kr:finma}. Here we select one example related to our focus when looking at data under the presence of moderate heavy tail ({\it i.e.} $\a>2$), but will draw conclusions based on all the results. \\


\n Let us consider the example of $\a=5/2$. \\


\vspace{2ex}

 
\hspace{-7ex} \begin{minipage}{0.45\linewidth}
\centering
\begin{tabular}{c|c|c|c|c}
$n=52$ & {\tiny Simul} &  {\tiny  CLT}  & {\tiny Max}  & {\tiny Normex}  \\
$q$ & $z_q$ & $z_q^{(2)}$ & $ z_q^{(3)}$ &  $z_q^{(5)}$  \\
  &  & $ \delta^{(1)} $ (\%) & $ \delta^{(3)}  (\%) $  &  $\delta^{(5)}  (\%) $ \\
\hline\hline 
95\% &  103.23 & 104.35  &  102.60 &  103.17  \\
 & &1.08  & -0.61 & {\bf -0.06} \\
\hline
99\% & 119.08 & 111.67   & 117.25 &  119.11   \\
 & & -6.22 & -1.54 &  {\bf 0.03} \\
\hline
99.5\% & 128.66  & 114.35   & 127.07 & 131.5   \\
 & & -11.12 & -1.24 &  2.21  \\
\end{tabular}
\end{minipage}
\hfill
\begin{minipage}{0.45\linewidth}
\centering
\begin{tabular}{c|c|c|c|c}
$n=100$ & {\tiny Simul} &  {\tiny  CLT}  & {\tiny Max}  & {\tiny Normex}  \\
$q$ & $z_q$ & $z_q^{(2)}$ & $ z_q^{(3)}$ &  $z_q^{(5)}$  \\
  &  & $ \delta^{(1)} $ (\%) & $ \delta^{(3)}  (\%) $  &  $\delta^{(5)}  (\%) $ \\
\hline\hline 
95\% &  189.98 & 191.19   &  187.37 &   189.84  \\
 & & 0.63  & -1.38 & {\bf -0.07}  \\
\hline
99\% & 210.54  & 201.35   & 206.40 &  209.98    \\
 & & -4.36  & -1.96 & {\bf -0.27} \\
\hline
99.5\% & 222.73  & 205.06   &  219.14 &  223.77  \\
 & & -7.93 & -1.61 &  {\bf 0.47} \\
\end{tabular}
\end{minipage}

\vspace{2ex}

\hspace{-7ex} \begin{minipage}{0.45\linewidth}
\centering

\begin{tabular}{c|c|c|c|c}
$n=250$ & {\tiny Simul} &  {\tiny CLT} & {\tiny Max} & {\tiny Normex} \\
$q$ & $z_q$ & $z_q^{(2)}$ &  $ z_q^{(3)}$ &    $z_q^{(5)}$  \\
 &  & $\delta^{(1)}$ (\%) & $ \delta^{(3)}$ (\%) &  $\delta^{(5)}$ (\%) \\
\hline\hline 
95\% & 454.76 & 455.44 & 446.53 & 453.92     \\
& & 0.17  & -1.81 &  {\bf -0.18}  \\
\hline
99\% & 484.48  & 471.5 & 473.99 & 483.27     \\
& & -2.68  & -2.17 &  {\bf -0.25} \\
\hline
99.5\% &  501.02 &  477.38 & 492.38 & 501.31    \\
& & -4.72  & -1.73 & {\bf 0.06}  \\
\end{tabular}
\end{minipage}
\hfill
\begin{minipage}{0.45\linewidth}
\centering
\begin{tabular}{c|c|c|c|c}
$n=500$  & {\tiny Simul} &  {\tiny CLT} & {\tiny Max} &  {\tiny Normex}  \\
$q$ & $z_q$ & $z_q^{(2)}$ &  $ z_q^{(3)}$ &    $z_q^{(5)}$  \\
 &  & $\delta^{(1)}$ (\%) & $ \delta^{(3)}$ (\%) &  $\delta^{(5)}$ (\%) \\
\hline\hline 
95\% & 888.00  & 888.16 & 872.74 &  886.07   \\
& & 0.02  & -1.72 & -0.22 \\
\hline
99\% & 928.80 & 910.88 & 908.97 &  925.19    \\
& & -1.93 & -2.14 & {\bf -0.39}  \\
\hline
99.5\% & 950.90 & 919.19 &  933.23 & 948.31    \\
& & -3.33  & -1.86 & {\bf -0.27} \\
\end{tabular}
\end{minipage}

\subsubsection{Discussion of the results}

\vspace*{2ex}
\begin{itemize}
\item Those numerical results are subject to numerical errors due to the finite sample of simulation of the theoretical value, as well as the choice of random generators, but the most important reason for numerical error of our methods resides in the convergence of the integration methods. Thus, one should read the results, even if reported with many significant digits, to a confidence we estimate to be around 0.1\%  
\item  Concerning Normex, we find out that:
\begin{itemize}
\item[-] the accuracy of the results appears more or less independent of the sample size $n$, which is the major advantage of our method when dealing with the issue of aggregation
\item[-] for $\a >2$, it always gives sharp results (error less than 0.5\% and often extremely close); for most of them, the estimation is indiscernible from the true value, obviously better than the ones obtained with the other methods 
\item[-] for $\a\le 2$, the results for the most extreme quantile are less satisfactory than expected. We attribute this to a numerical instability in the integration procedure used in R. Indeed, for very large quantiles ($\ge 99.5$\%), the convergence of the integral seems a bit more unstable (due to the use of the package Vegas in R), which may explain why the accuracy decreases a bit, and may sometimes be less than with the max method. We plan to explore this problem further.
\end{itemize}
\item The max method overestimates for $\a <2$ and underestimates for $\a\ge 2$; it improves a bit  for higher quantiles and $\a\le 2$. It is a method that practitioners  should think about, because it is very simple to use and gives already a first good approximation for the VaR (as the CLT does for the mean)
\item The GCLT method ($\a<2$) overestimates the quantiles but improves with higher quantiles and when $n$ increases
\item  Concerning the CLT method, we find out that:
\begin{itemize}
\item[-] the higher the quantile, the higher the underestimation; it improves slightly when $n$ increases, as expected
\item[-] the smaller $\a$, the larger the underestimation 
\item[-] for $\a\ge 2$,  the VaR evaluated with the normal approximation is always lower than the VaR evaluated via Normex. The lower $n$ and $\a$, the higher the difference
\item[-] the difference between the VaR estimated by the CLT and the one estimated with Normex, appears large for relatively small $n$, with a relative error going up to 13\%, and decreases when $n$ becomes larger  
\end{itemize}
\item We have concentrated our study on the VaR risk measure because it is the one used in solvency regulations both for banks and insurances. However, the Expected Shortfall, which is the only coherent measure in presence of fat tails, would be more appropriate for measuring the risk of the companies. The difference between the risk measure estimated by the CLT and the one estimated with Normex would certainly be  much larger than what we obtain with the VaR, when the risk is measured with the Expected Shortfall, pleading for using this measure in presence of fat tails.  
\end{itemize}
 
\section*{Conclusion}

\n The main motivation of this study was to propose a sharp approximation of the entire distribution of aggregated risks under the presence of fat tails, in view of application on financial or insurance data. In particular  the aim was to obtain the most accurate  evaluations of risk measures.
There was still mathematically a missing 'brick' in the literature on the behavior of the sum of iid rv's with a moderate heavy tail, for which the CLT applies but with a slow convergence for the mean behavior and certainly does not provide satisfactory approximation for the tail. Our study fills up this gap, by looking at an appropriate limit distribution.

\n After reviewing the existing methods,  we built  {\it Normex}, a method mixing a limit normal distribution via the CLT  and the exact distribution of a small number (defined according to the range of $\a$ and the choice of the number of existing moments of order $p$) of the largest order statistics.

\n In this study, Normex has been proved, theoretically as well as numerically, to deliver a sharp approximation of the true distribution, for any sample size $n$ and for any positive tail index $\alpha$, and generally better than what existing methods provide. 

\n An advantage of Normex is its generality. Indeed, trimming the total sum by taking away extremes having infinite moments (of order $p\ge 3$) is always possible and allows to better approximate the distribution of the trimmed sum with a normal one (via the CLT). Moreover, fitting a normal distribution for the mean behavior can apply, not only for the Pareto distribution, but for any underlying distribution, without having to know about it, whereas for the extreme behavior, we pointed out that a Pareto type is standard in this context.

\n Normex could also be used from another point of view. We could apply it for a type of inverse problem, to find out a range for the tail index $\a$ when fitting this explicit mixed distribution to the empirical one. 
Other approaches have been proposed to estimate the tail index, which may be classified into two classes, supervised procedures in which the threshold to estimate the tail is chosen according to the problem (as e.g. the MEP~(\cite{da:sm}), Hill~(\cite{hill}), or QQ~(\cite{kr:sid}) methods) and unsupervised ones, where the threshold is algorithmically determined  (as e.g. in \cite{ca:be}, \cite{nehla}). Normex would be a new unsupervised approach, since the $k$ is chosen algorithmically for a range of $\a$.

\n Other perspectives concern the application of this study to real data,  as well as its extension to the dependent case for which some interesting recent results on stable limits for sums of dependent infinite variance r.v. from Bartkiewicz et al. (see \cite{bar:jmw}) and Large Deviation Principles from Mikosch et al. (see \cite{mik:w}) may be useful.\\

\noindent {\small {\it Acknowledgement.} This work has been carried out during the author's internship at the Swiss Financial Market Supervisory Authority (FINMA); the author would like to thank both Hansj\"org Furrer and FINMA for their hospitality and rich environment, and for bringing to her attention  this interesting practical issue. Warm thanks also to Michel Dacorogna, for stimulating and interesting discussions on this study.} \\~

\end{document}